\documentclass[aps,pra,showpacs,twoside,10pt,floatfix,nofootinbib,longbibliography,twocolumn]{revtex4-1}
\usepackage[colorlinks=true, citecolor=red, urlcolor=blue ]{hyperref}
\usepackage{epsfig,newlfont,amssymb,amsfonts,amsmath,bm,subfigure,palatino,mathtools,amsthm,braket,soul,enumitem,color,times,comment,geometry,xcolor}
\usepackage[normalem]{ulem}
\newcommand{\propose}[1]{{\noindent\textbf{$\blacksquare$ Proposition #1.}}}

\begin{document}

\title{Block entanglement bounds distribution of regionally localized entanglement}
\author{Jithin G. Krishnan$^1$, Aditi Sen(De)$^2$, Amit Kumar Pal$^1$}
\affiliation{$^1$Department of Physics, Indian Institute of Technology Palakkad, Palakkad 678 623, India}
\affiliation{$^2$Harish-Chandra Research Institute, A CI of Homi Bhabha National Institute, Chhatnag Road, Jhunsi, Allahabad - 211019, India}

\begin{abstract}

In quantum networks, eliminating connections between nodes is crucial to mitigate the effects of decoherence, often achieved by performing measurements on nodes that are idle, or vulnerable to noise. To characterize the entanglement content of the resulting smaller network, we introduce the notion of  ``regionally localized entanglement", defined as the average entanglement  concentrated over a two-qubit region in a multi-qubit system. Hence, the total regionally localized entanglement can be obtained by considering  all two-qubit regions sharing a common qubit, referred to as the ``hub".  We prove that the total regionally localized entanglement corresponding to a specific hub is bounded above and below via the  localizable block entanglement shared between the hub and the rest of the multi-qubit system for a number of paradigmatic pure quantum states, including permutation-symmetric states and arbitrary superposition of states from a specific magnetization sector. Numerical simulations confirm that the bounds for permutation-symmetric pure states remain valid even for Haar-uniformly generated  pure states, and when each of the qubits is sent through local phase-flip channels of Markovian and non-Markovian types, except when the system-size is small. On the other hand, arbitrary states from a particular magnetization sector yield bounds on total regionally localized entanglement that are distinct from the permutation-symmetric states, highlighting the structurally unique entanglement properties of the former.
\end{abstract}

\maketitle

\section{Introduction}
\label{sec:intro}

Quantum networks~\cite{Kimble2008,Luo2024} require the distribution of entanglement across multiple nodes while ensuring that the generated entanglement remains resilient against decoherence~\cite{breuer2002,Lidar2013}. In this context, it is often advantageous to eliminate certain connections, or nodes that are ineffective, i.e, either idle, or particularly susceptible to environmental noise. There are two principal approaches to such elimination -- either by discarding (i.e., mathematically tracing out) the ineffective nodes, or by actively performing local operations on them. Starting from an entangled state spread over the entire network, tracing out the ineffective nodes typically yields a mixed state with reduced entanglement between the remaining operational nodes (see~\cite{horodecki2009,das2016,dechiara2018} and the references therein). In contrast, local measurements on the ineffective nodes, followed by appropriate classical communication and postselection, can enhance entanglement over the unmeasured nodes. This observation has led to the development of concepts like the entanglement of assistence (EoA)~\cite{divincenzo1998,Laustsen2003,Gour2006} and localizable entanglement (LE)~\cite{verstraete2004,verstraete2004a,popp2005,sadhukhan2017}, which quantifies the maximum average entanglement that can be concentrated among a subset of nodes through local measurements on the rest. This notion plays a significant role in the analysis of multipartite entangled states such as Greenberger Horne Zeilinger (GHZ) states~\cite{greenberger1989}, graph states~\cite{hein2006}, and stabilizer states in quantum error correcting codes~\cite{gottesman2010,Girvin2023}. Despite the computational challenge due to the optimization inherent in its definition and its dependence on the choice of the entanglement measure, the latter approach has been explored extensively for characterizing entanglement in the GHZ and GHZ-like states~\cite{hein2006,amaro2018,amaro2020a,HK2023} in absence as well as presence of noise, in exploring quantum correlations present in interacting quantum spin models~\cite{verstraete2004,verstraete2004a,popp2005,jin2004,skrovseth2009,smacchia2011,montes2012,sadhukhan2017,HK2022,HK2023,HK2025,Krishnan2025}, and in developing ideas like entanglement percolation through quantum networks~\cite{acin2007}.

Given the role of entanglement as a resource~\cite{horodecki2009} in quantum information theoretic protocols~\cite{bennett1993,bouwmeester1997,bennett1992,mattle1996,sende2010,ekert1991,jennewein2000,raussendorf2001,raussendorf2003,briegel2009},
characterizing the entanglement content within a quantum network is of paramount importance. However, this task becomes particularly challenging in realistic scenarios involving noise~\cite{breuer2002,Lidar2013}, as quantification of  entanglement in general multipartite mixed states remains elusive~\cite{horodecki2009,das2016,dechiara2018}. In this context, understanding the fundamental limits on the extent with which a particular subsystem can be entangled with others in a multipartite quantum state becomes essential. To address this, the notion of monogamy of entanglement~\cite{ckw2000,Terhal2004,Christandl2004,Osborne2006,Kim2012a} has been introduced, establishing constraints on how entanglement can be shared among subsystems, where it is primarily calculated on the mixed states of subsystems obtained by tracing out the rest of the subsystems. However,  explorations of appropriate bounds on the distribution of the average entanglement quantified, in the same spirit as in the cases of the EoA~\cite{divincenzo1998,Laustsen2003,Gour2006} and the LE~\cite{verstraete2004,verstraete2004a,popp2005,sadhukhan2017}, have been relatively sparse. A relation dual to the monogamy of entanglement~\cite{ckw2000,Terhal2004,Christandl2004,Osborne2006,Kim2012a} has been established for EoA~\cite{Gour2005,Gour_2007,Buscemi_2009,Kim_2012,Kim2012a,Kim2014,Kim2018,Guo2018,Yang2019,Jin2019,Jin2019b,shen2024}. Also bounds on the EoA~\cite{pollock2021} and the LE~\cite{Krishnan2023} over a given bipartition of a chosen subsystem in a multi-qubit system have been derived in terms of the bipartite entanglement content between the chosen subsystem and the rest of the system in the original state. However, despite these attempts, a complete understanding of the  possible meaningful bounds on the shareable average entanglement localized between different subsystems in a multipartite system is still lacking.

\begin{figure*}
    \includegraphics[width=0.8\linewidth]{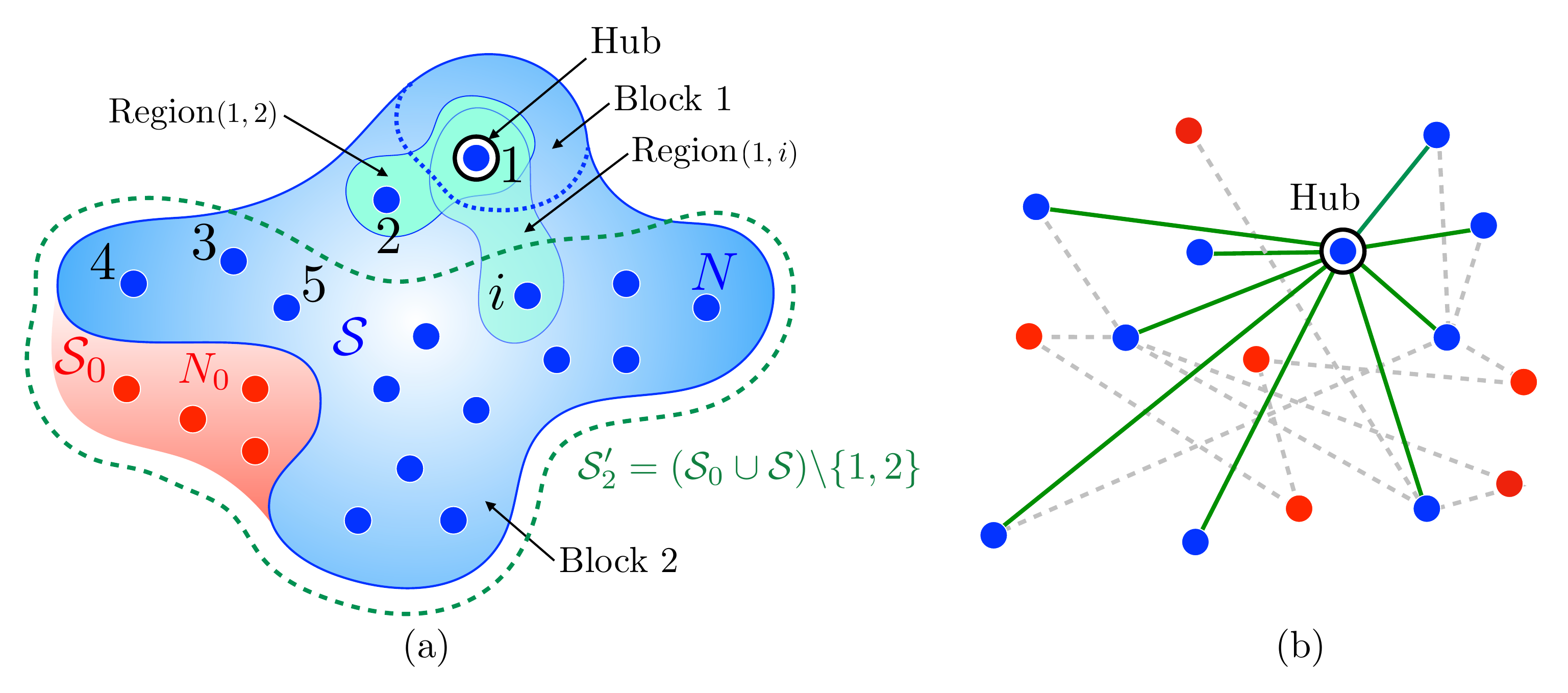}
    \caption{
    (a) \textbf{Schematic representation of local regions and blocks in a multi-qubit system.} A system of $M=N_0+N$ qubits is divided into two parts, $\mathcal{S}_0$ and $\mathcal{S}$, where the set $\mathcal{S}_0$ $(\mathcal{S})$ contains  $N_0$ ($N$) qubits, and the qubits in $\mathcal{S}$ are labeled as $1,2,\cdots,N$, with the qubit $1$ as the hub. The subsystem $\mathcal{S}$ is divided into two blocks -- one block, referred to as the block 1, containing only the hub, while the other block, referred to as the block 2, containing the rest $N-1$ qubits. Each qubit pair $(1,i)$ including the hub qubit $1$ and a qubit $i$ ($i=2,3,\cdots,N$) constitutes a \emph{local} region in $\mathcal{S}$, demonstrated for the qubit pairs $(1,2)$ and $(1,i)$. The set $\mathcal{S}^\prime_{i}$ includes all qubits in the set $\mathcal{S}_0\cup\mathcal{S}$ except the qubits $1$ and $i$, which is indicated as $(\mathcal{S}_0\cup\mathcal{S})\backslash\{1,i\}$. While studying the distribution of localizable entanglement among parties in \(\mathcal{S}\), we perform measurements on $(\mathcal{S}_0\cup\mathcal{S})\backslash\{1,i\}$  (\(i=2, \ldots, N\)) to compute \(\mathcal{F}_i(\rho)\) on the region constituted by the qubits $1$ and $i$. To compare this with block entanglement content in \(\rho\),  local measurements are performed on \(\mathcal{S}_0\), leading to \(\mathcal{E}_{\mathcal{S}}(\rho)\), where entanglement in each post-measured state is computed over the bipartition between the hub qubit (block 1) and the rest of the qubits (block 2). (b) \textbf{An example of a quantum network with a hub directly connected to a large number of nodes}. The hub together with the nodes connected to it forms the subsystem $\mathcal{S}$, while all other nodes that are not directly connected to the hub form the subsystem $\mathcal{S}_0$. This resembles the setup described in (a), considering each node to be a qubit.}  
    \label{fig:partition}
\end{figure*}

Towards bridging this gap, we investigate whether and how the entanglement localized over a ``region" involving two parties of a multipartite quantum system is bounded.  More specifically, in a multi-qubit system,  we consider the LE over the local regions constituted by qubit-pairs with one qubit, called the ``hub" from now onward, as common in all such regions (see Fig.~\ref{fig:partition}(a)), and refer to this LE as the ``regionally localized entanglement" (RLE). We determine the bounds of the total RLE corresponding to a hub imposed due to (a) the  localized entanglement between the hub and the rest of the subsystems, referred to as the ``block localized entanglement" (BLE),   and (b) the entanglement between the hub and the rest, calculated using the reduced state of either of the subsystem, and referred to as the ``block entanglement" (BE) (see Fig.~\ref{fig:partition}(a)). Our construction is  relevant in the practical and contemporary problem of concentrating entanglement over regions of useful nodes in large quantum networks for performing quantum tasks, where one node, forming a hub, is common to all such regions, as our results provide practical bounds on the distribution of the LE over different links connected to the hub of the large quantum network (see Fig.~\ref{fig:partition}(b)).

We prove upper bounds in terms of the BLE and the BE on the total RLE  for paradigmatic classes of multi-qubit pure states, namely, the permutation-symmetric states including the GHZ state~\cite{greenberger1989},  the generalized GHZ (gGHZ) states~\cite{dur2000}, the W state~\cite{dur2000,sende2003}, the Dicke states~\cite{dicke1954} and their superpositions~\cite{sadhukhan2017}, along with states within a specific magnetization sector that break permutation symmetry, such as  the generalized  W (gW) states~\cite{dur2000}.  
On the other hand, for states with specific magnetization but without permutation symmetry,  especially the gW states with a single excitation, we derive both lower and upper bounds via the BLE and the BE, which turn out to be different than the one obtained for the permutation-symmetric states. We further consider Haar-uniformly~\cite{bengtsson2017} generated pure arbitrary multi-qubit states, and numerically demonstrate the applicability of the derived  bounds. In order to see whether noise plays a role in violating these bounds,  each of the qubits in the multi-qubit system is subjected to a Markovian and a non-Markovian phase-flip channel,  and we find that a violation of the derived bounds takes place only when the number of qubits in the system is small. With large system-sizes, the derived bounds are found to hold in the case of both Markovian and non-Markovian phase-flip noise.

The rest of the paper is organized as follows. Sec.~\ref{sec:definitions} sets the stage, describing the concepts of  the block-wise and the regional LE. The analytical calculation of the bounds for paradigmatic pure states as well as the testing of the derived bounds for arbitrary multi-qubit states are discussed in Sec.~\ref{sec:pure_states}. The influence of the presence of local Markovian and non-Markovian phase-flip noise on the validity of the derived bounds are explored in Sec.~\ref{sec:local_dephasing}. Sec.~\ref{sec:conclude} contains the conclusive remarks and outlook.

\section{Introducing the notion of regionally localized entanglement}
\label{sec:definitions}

Consider an $M$-qubit system described by the state $\rho$, constituted of a 
partition $\mathcal{S}_0$ of $N_0$ qubits, and another partition $\mathcal{S}$ of $N$-qubits, 
where we label the qubits in $\mathcal{S}$ as $1,2,\cdots,N$ (see Fig.~\ref{fig:partition}). 
Performing  projection measurements on all $\mathcal{S}_0$ qubits leads to a post-measured ensemble $\left\{p_k,\varrho_{\mathcal{S}}^k=\text{Tr}_{\mathcal{S}_0}\left[\varrho^k\right]\right\}$, with  
\begin{eqnarray}\label{eq:post_measured_state}
    \varrho^k&=&\frac{1}{p_k}\left[\left(I_{\mathcal{S}}\otimes P_{\mathcal{S}_0}^k\right)\rho\left(I_{\mathcal{S}}\otimes P_{\mathcal{S}_0}^k\right)\right]
\end{eqnarray}
being the post-measurement state on the $N$-qubit system corresponding to the measurement outcome $k$, occurring with the probability 
\begin{eqnarray}
    p_k=\text{Tr}\left[\left(I_{\mathcal{S}}\otimes P_{\mathcal{S}_0}^k\right)\rho\left(I_{\mathcal{S}}\otimes P_{\mathcal{S}_0}^k\right)\right]. 
\end{eqnarray}
Here, $I_{\mathcal{S}}$ is the identity operator on the Hilbert space of $\mathcal{S}$, and $P_{\mathcal{S}_0}^k=\ket{b_k}\bra{b_k}$ is the rank-$1$ projection operator corresponding to the measurement outcome $k$, $\{\ket{b_k};k=0,1,\cdots,2^{N_0}-1\}$ being the measurement-basis (which can, in principle, be an \emph{entangled} basis). 

\paragraph{Block localizable entanglement.} Choosing an appropriate bipartite or multipartite entanglement measure $E$, one defines the localizable entanglement  (LE) as the maximum of the average entanglement concentrated on the set $\mathcal{S}$ of the qubits via projection measurement on $\mathcal{S}_0$, given by 
\begin{eqnarray}\label{eq:lqc}
    \mathcal{E}_{\mathcal{S}}(\rho)=\max_{\mathcal{M}_0}\sum_{k}p_k E(\varrho_{\mathcal{S}}^k),
\end{eqnarray}
where the maximization is performed over the set of all possible projection measurements, $\mathcal{M}_0$, on $\mathcal{S}_0$. Note that with the choice of a \emph{convex} entanglement measure $E$, $\mathcal{E}_{\mathcal{S}} (\rho)\geq E(\rho_{\mathcal{S}})$, 
where $\rho_{\mathcal{S}}=\text{Tr}_{\mathcal{S}_0}\left[\rho\right]$. In this paper, we concentrate over bipartite entanglement measures, specifically negativity~\cite{peres1996,horodecki1996,zyczkowski1998,vidal2002,lee2000} (see Appendix~\ref{app:negativity} for the definition),  which, unless otherwise stated, is always computed over a bipartition of $\mathcal{S}$ into two blocks, where one block consists of only one qubit, referred to as the hub, and the other block hosts the rest of the useful qubits in the network.  Without loss of generality, the hub qubit is considered to be the qubit $1$, such that $\mathcal{E}_{\mathcal{S}}(\rho)$ is the \emph{block localizable entanglement} (BLE) shared between the hub and the rest of the qubits in $\mathcal{S}$. In the special case of $\mathcal{S}_0=\emptyset$, bipartite entanglement between the hub qubit $1$ and the rest of the qubits in $\mathcal{S}$ together, as quantified by negativity~\cite{peres1996,horodecki1996,zyczkowski1998,vidal2002,lee2000}, is given by the \emph{block entanglement} (BE) $E(\rho=\rho_\mathcal{S})$.

\paragraph{Regionally localized entanglement.} Let us now consider the set $\{\mathcal{F}_i(\rho)\}$ of the values of localizable entanglement computed over all two-qubit regions $(1,i)$ ($i=2,3,\cdots,N$), where $\mathcal{F}_i(\rho)$ is referred to as the \emph{regionally localized entanglement} (RLE). Here, the qubit $1$ is again chosen as the hub. The calculation of $\mathcal{F}_i(\rho)$ involves measuring the subsystem $\mathcal{S}^\prime_{i}=(\mathcal{S}_0\cup\mathcal{S})\backslash\{1,i\}$, i.e., all $M-2$ qubits barring qubits $1$ and $i$ (see Fig.~\ref{fig:partition}(a) for the example of $\mathcal{S}^\prime_i$), such that  
\begin{eqnarray}\label{eq:two_qubit_entanglements}
\mathcal{F}_i(\rho)&=&\max_{\mathcal{M}^\prime_{i}}\sum_{k}p_k E(\varrho_{1i}^k).
\end{eqnarray}    
Here, $\mathcal{M}^\prime_{i}$ is the set of all possible projection measurements on the set $\mathcal{S}^\prime_{i}$,  $\varrho_{1i}^k=\text{Tr}_{\mathcal{S}^\prime_i}\left[\varrho^k\right]$, and $E$ is the same bipartite entanglement measure used for calculating $\mathcal{E}_{\mathcal{S}}(\rho)$. One can further calculate the total RLE, given by    
\begin{eqnarray}
\mathcal{F}_{\mathcal{S}}(\rho)=\sum_{i=2}^N\mathcal{F}_{i}(\rho),
\label{eq:multiparty_in_terms_of_two_qubit}
\end{eqnarray}
corresponding to the hub $1$. 

In this paper, we explore the interplay between  $E(\rho)$, $\mathcal{E}_{\mathcal{S}}(\rho)$, and $\mathcal{F}_{\mathcal{S}}(\rho)$ in the case of large arbitrary multi-qubit states, pure as well as mixed, which is challenging due to the increase in the number of state parameters with $M$ as well as the optimizations involved in the definitions of $\mathcal{E}_{\mathcal{S}}(\rho)$, and $\mathcal{F}_{\mathcal{S}}(\rho)$.  In this context, we point towards the idea of monogamy of entanglement~\cite{ckw2000,Terhal2004,Christandl2004,Osborne2006,Kim2012a} (see Appendix~\ref{app:monogamy}), where similar line of investigation is followed with partial trace-based quantification of entanglement over qubit pairs. Note further that  the EoA~\cite{divincenzo1998,Laustsen2003,Gour2006}, of which the LE~\cite{verstraete2004,verstraete2004a,popp2005,sadhukhan2017} is a special case, has also been explored~\cite{Gour2005,Gour_2007,Buscemi_2009,Kim_2012,Kim2012a,Kim2014,Kim2018,Guo2018,Yang2019,Jin2019,Jin2019b,shen2024} under the glass of the monogamy of entanglement. However, the fact that both $\mathcal{E}_{\mathcal{S}}(\rho)$ and $\mathcal{F}_{\mathcal{S}}(\rho)$ are upper bounded by the EoA (see Appendix~\ref{app:eoa}) makes the exploration of the relations between $E(\rho)$, $\mathcal{E}_{\mathcal{S}}(\rho)$, and $\mathcal{F}_{\mathcal{S}}(\rho)$ non-trivial.

To compute $\mathcal{E}_{\mathcal{S}}(\rho)$ and $\mathcal{F}_{\mathcal{S}}(\rho)$,  we consider independent rank-$1$ \emph{local} projection measurements on each qubit in the set, say, $\mathcal{S}_0$.  The corresponding projectors are given by  $P_{\mathcal{S}_0}^k=\otimes_{i\in \mathcal{S}_0}P_i^{k_i}$, with $k\equiv k_1k_2\cdots k_{N_0}$.  Here, $P_i^{k_i}$ is the projection operator on the qubit $i\in \mathcal{S}_0$ corresponding to the measurement outcome $k_i$, such that $P_i^{k_i}=\ket{b_{k_i}}\bra{b_{k_i}}$, $k_i=0,1$, and 
\begin{eqnarray}
    \ket{b_0} &=& \cos\frac{\theta_i}{2}\ket{0_i}+\text{e}^{\text{i}\phi_i}\sin\frac{\theta_i}{2}\ket{1_i},\nonumber\\
    \ket{b_1} &=& \sin\frac{\theta_i}{2}\ket{0_i}-\text{e}^{\text{i}\phi_i}\cos\frac{\theta_i}{2}\ket{1_i},   
\end{eqnarray}
are the measurement-basis in the qubit Hilbert space, $\theta_i,\phi_i\in\mathbb{R}$, $\theta_i\in[0,\pi]$, $\phi_i\in[0,2\pi]$. The maximization over $\mathcal{M}_0$ is achieved via a maximization over the $2N_0$ real parameters $\{(\theta_i,\phi_i):i\in \mathcal{S}_0\}$. Similar construction works for measurements on qubits belonging to $\mathcal{S}^\prime_i$ and the corresponding maximization over the set $\mathcal{M}^\prime_i$ involves optimization over $2(M-2)$ real parameters.

\section{Boundaries of the regionally localized entanglement}
\label{sec:pure_states}

Let us now determine the bounds on the total RLE in paradigmatic families of multi-qubit pure states, using the BLE and the BE.  Consider the $2^N$-dimensional complex Hilbert space $\mathcal{H}$ corresponding to the $N$-qubit system with $N\geq 3$, spanned by the orthonormal computational basis $\{\ket{b_l},l=0,1,\cdots,2^N-1\}$, where $\ket{b_l}=\otimes_{i=1}^N\ket{b_{i,l}}$, with $\ket{b_{i,l}}$ defined as $\sigma^z_i\ket{b_{i,l}}=b_{i,l}\ket{b_{i,l}}$, such that the qubit $i$ ($i=1,2,\cdots,N$) in the state $\ket{b_l}$ has a \emph{magnetization}  $b_{i,l}=\pm 1$. Each basis state $\ket{b_l}$ can be labeled by the \emph{total magnetization} $m$, as $\ket{b_l^m}$, defined by $\sum_{i=1}^N\sigma^z_i\ket{b_l^m}=m\ket{b_l^m}$, with $m=2n-N$, where $n$ is the number of qubits in $\ket{b_l^m}$ having $b_{i,l}=+1$. The magnetizations $m$ assumes $N+1$ possible values, given by $m=-N,-N+2,\cdots,N-2,N$, where each value has a multiplicity $\mathcal{N}_m=\genfrac(){0pt}{1}{N}{n}$, and corresponds to a subspace $\mathcal{H}_m$ of the full Hilbert space, spanned by the states $\{\ket{b_l^m}\}$. A generic state in $\mathcal{H}_m$ is defined as 
\begin{eqnarray}
    \ket{\psi_m}=\sum_{k=1}^{\mathcal{N}_m}c_k\ket{b^m_{l,k}},
    \label{eq:magnetization_sector_generic_state}
\end{eqnarray}
with $c_k$ being complex such that $\sum_{k=1}^{\mathcal{N}_m}|c_k|^2=1$, where the index $k$ takes care of the multiplicity of the total magnetization.  

\subsection{Permutation-symmetric states}
\label{subsec:permutation_symmetry_1}

A state is said to be permutation-symmetric if it remains invariant under all permutations of any number of parties. By imposing this condition on $\ket{\psi_m}$, we obtain the $N$-qubit Dicke state with $n$ excitations, written in the computational basis as 
\begin{eqnarray}
    \ket{D_m}=\frac{1}{\sqrt{\genfrac(){0pt}{1}{N}{n}}}\sum_{i}\mathcal{P}_i\left[\ket{0}^{n}\ket{1}^{N-n}\right],
    \label{eq:dicke_computational_basis}
\end{eqnarray}
where $\{\mathcal{P}_i\}$ is the set of all possible permutations of $N$-qubits, $n$ of which are in the state $\ket{0}$ (excited state), and $N-n$ in $\ket{1}$ (ground state). 
On the other hand, a permutation-symmetric state can also be defined across the different magnetization subspaces $\mathcal{H}_m$, as
\begin{eqnarray}
    \ket{\Psi}=\sum_{m=-N}^N c_m\ket{D_m},
    \label{eq:ghz_type}
\end{eqnarray}
with $c_m$ complex, and $\sum_m|c_m|^2=1$.

We first consider the situation of $\mathcal{S}_0=\emptyset$, in which the following proposition involving $E(\rho)$ and $\mathcal{F}_{\mathcal{S}}(\rho)$ holds.

\begin{figure}
    \centering
    \includegraphics[width=0.9\linewidth]{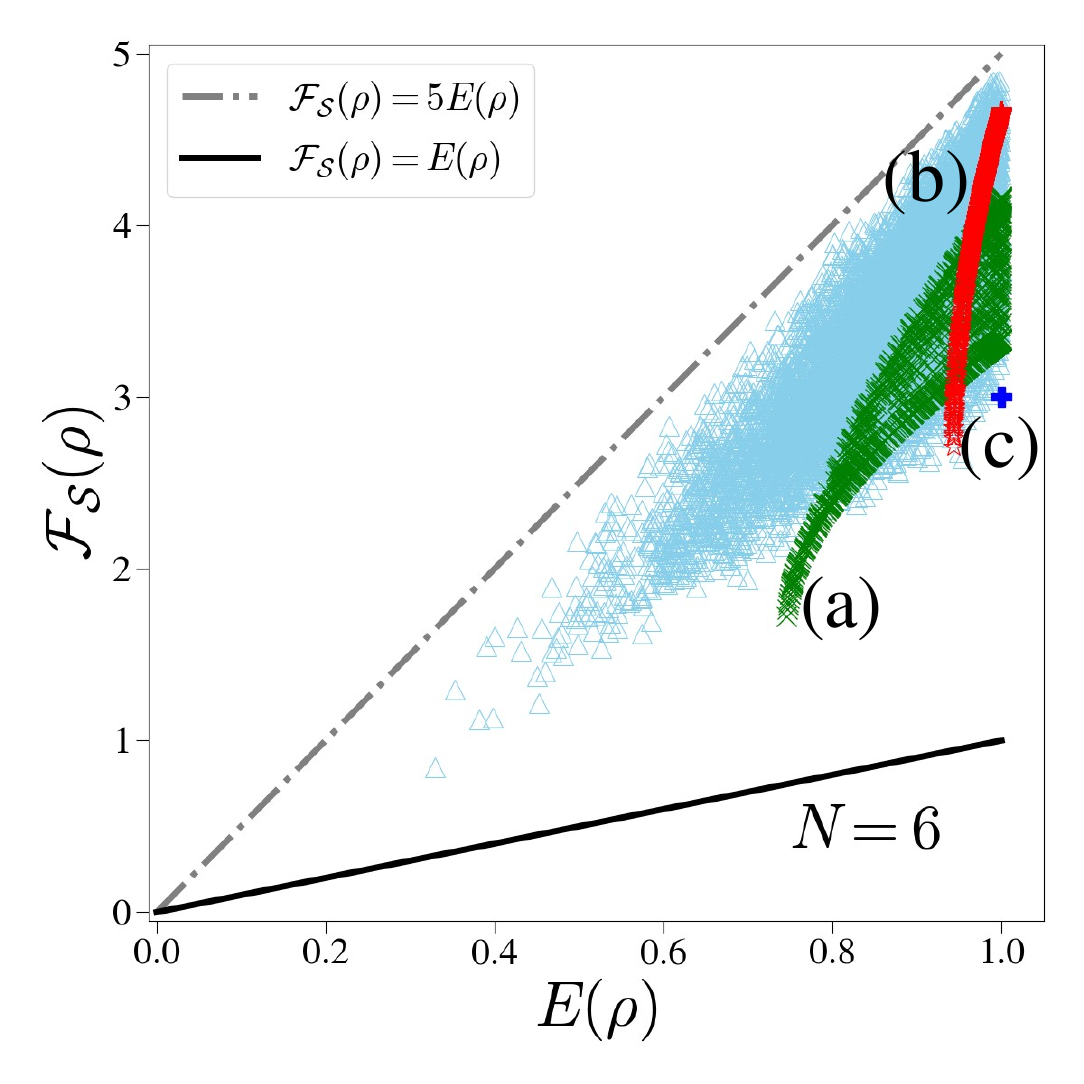}
    \caption{\textbf{Map of six-qubit permutation-symmetric states on the $\left(E(\rho),\mathcal{F}_{\mathcal{S}}(\rho)\right)$-plane.} Scatter plot of $\mathcal{F}_{\mathcal{S}}(\rho)$ (ordinate) against $E(\rho)$ (abscissa) for $10^6$ randomly generated states of the form $\ket{\Psi}$ (Eq.~(\ref{eq:ghz_type})) with $N=6$. The dash-dotted and continuous lines represent the lines,  $\mathcal{F}_{\mathcal{S}}(\rho)=5E(\rho)$ and $\mathcal{F}_{\mathcal{S}}(\rho)=E(\rho)$ respectively (see Proposition 1 and the following  discussion). The single parameter  states (a) $\ket{\Psi_{(+4,-4)}}$ and (b) $\ket{\Psi_{(+2,-2)}}$ (see Eq.~(\ref{eq:symmetric_state_magnetization_pair})), along with (c) the $6$-qubit Dicke state $\ket{D_0}$ with vanishing magnetization (see Eq.~(\ref{eq:dicke_computational_basis}))    are also marked. All quantities plotted are dimensionless.}
    \label{fig:s0_empty_symmetric}
\end{figure}  

\propose{1} \emph{For an $N$-qubit system described by a permutation-symmetric pure state,}
\begin{eqnarray}
\mathcal{F}_{\mathcal{S}}(\rho)\leq (N-1)E(\rho).
\label{eq:proposition_1}
\end{eqnarray}

\begin{proof}
    Let us consider the qubit-pair $\{1,2\}$ from the $N$-qubit system, $\mathcal{S}$. Monotonicity of entanglement dictates \begin{eqnarray}
        \mathcal{F}_2(\rho)&\leq&\min\left\{E_{1:23\cdots N}(\rho),E_{2:13\cdots N}(\rho)\right\},\nonumber\\
        &=& E_{1:23\cdots N}(\rho)=E(\rho),
    \end{eqnarray} 
    due to permutation symmetry,  which further implies $\mathcal{F}_i(\rho)\leq E(\rho)$ $\forall i\in\{2,3,\cdots,N\}$.  This, along with (\ref{eq:multiparty_in_terms_of_two_qubit}) leads to $\mathcal{F}_{\mathcal{S}}(\rho)\leq (N-1)E(\rho)$.
\end{proof}

For the $N$-qubit Dicke state with single excitation ($n=1$ in Eq.~(\ref{eq:dicke_computational_basis})), also known as the $N$-qubit W state $\ket{W}$~\cite{dur2000,sende2003}, $\mathcal{F}_{i}(\rho)=2 /N$ $\forall i$ with $\sigma^z$ being the optimal basis~\cite{Krishnan2023}, leading to $\mathcal{F}_{\mathcal{S}}(\rho)=\sqrt{N-1}E(\rho)$, while for Dicke states with arbitrary $N$ and $n$ ($n>1$)~\cite{Krishnan2023,chen2020},
\begin{eqnarray}
    E(\rho)&=& 2\frac{\sqrt{n(N-n)}}{N},
\end{eqnarray}
and
\begin{eqnarray}
    \mathcal{F}_i(\rho)&=& \frac{2n(N-n)}{N(N-1)},
\end{eqnarray}
leading to
\begin{eqnarray}
    \mathcal{F}_{\mathcal{S}}(\rho)=\sqrt{n(N-n)}E(\rho).
\end{eqnarray}
Therefore, for Dicke states with arbitrary $N$ and $n (\geq 1)$, $E(\rho)<\mathcal{F}_{\mathcal{S}}(\rho)$ in addition to (\ref{eq:proposition_1}), i.e., $E(\rho)<\mathcal{F}_{\mathcal{S}}(\rho)\leq (N-1)E(\rho)$. Further, for states of the form (\ref{eq:ghz_type}), with a fixed $N$, one can consider $N/2$ ($(N+1)/2$) pairs $(m,-m)$ of non-zero magnetizations for even (odd) $N$, while an additional case $m=0$, corresponding to the state $\ket{D_0}$, exists for even $N$. Within the class of states given in Eq.~(\ref{eq:ghz_type}), one can define a family of single parameter states 
\begin{eqnarray}
    \ket{\Psi_{(m,-m)}}=c_m\ket{D_m}+c_{-m}\ket{D_{-m}}
    \label{eq:symmetric_state_magnetization_pair}
\end{eqnarray}
with $|c_m|^2+|c_{-m}|^2=1$, which include the $N$-qubit generalized GHZ (gGHZ) state as a special case (corresponding to the  $(N,-N)$ pair for both even and odd $N$), given by  
\begin{eqnarray}\label{eq:gGHZ_state}
    \ket{\text{gGHZ}}=c_0\ket{0}^{\otimes N}+c_1\ket{1}^{\otimes N},
\end{eqnarray}
in the computational basis, with renaming $c_0\leftrightarrow c_{N}$, $c_1\leftrightarrow c_{-N}$. For the $N$-qubit gGHZ states, $E(\rho)=\mathcal{F}_i(\rho)=2|c_0|\sqrt{1-|c_0|^2}$~\cite{Krishnan2023}, such that the upper bound  is reached, satisfying $E(\rho)<\mathcal{F}_{\mathcal{S}}(\rho)=E(\rho)(N-1)$. For the $N$-qubit GHZ state ($c_0=c_1=1/\sqrt{2}$),  $\mathcal{F}_{\mathcal{S}}(\rho)=N-1$ as $\mathcal{F}_{i}(\rho)=E(\rho)=1$ $\forall i$~\cite{divincenzo1998,popp2005,verstraete2004,verstraete2004a}.  For all other states of the form $\ket{\Psi_{(m,-m)}}$ with a fixed $N$ and $m\neq N$, our numerical analysis suggests  
$E(\rho)<\mathcal{F}_{S}(\rho)<(N-1)E(\rho)$. Similar findings are obtained for the states of the form in Eq.~(\ref{eq:ghz_type}) also.

When $\mathcal{S}_0\neq \emptyset$, there are $M=N_0+N$ qubits with $\mathcal{S}_0$ holding $N_0$ of them. In this case, for an arbitrary permutation-symmetric state of the form Eq.~(\ref{eq:ghz_type}), we write the following proposition involving $\mathcal{E}_{\mathcal{S}}$ and $\mathcal{F}_{\mathcal{S}}$.

\propose{2} \emph{For an $M$-qubit system described by a permutation-symmetric pure state,}
\begin{eqnarray}
 \mathcal{F}_{\mathcal{S}}(\rho)\leq (N-1)\mathcal{E}_\mathcal{S}(\rho).
\end{eqnarray}
\begin{proof}
    To prove this, we start with the calculation of $\mathcal{F}_i(\rho)$, and work with the optimal measurement basis corresponding to measurements on $\mathcal{S}_0\cup\mathcal{S}\backslash\{1,i\}$ that maximizes $\mathcal{F}_i(\rho)$.  Starting from a permutation-symmetric $M$-qubit pure state $\ket{\Psi}$, without any loss in generality, we  exploit the independence of the single-qubit projection measurements, and write the post-measured ensemble on $\mathcal{S}$ to be $\{p_{k_0},\ket{\Psi_{\mathcal{S}}^{k_0}}\}$, where $k_0$ denotes the measurement outcomes corresponding to measurements on $\mathcal{S}_0$. The average entanglement over the partition $1:\text{rest}$ in $\mathcal{S}$ is given by\footnote{Note that the chosen basis does not maximize this entanglement.} 
    \begin{eqnarray}
        \mathcal{E}^\prime_\mathcal{S}(\rho)=\sum_{k_0} p_{k_0} E_{1:23\cdots N}(\ket{\Psi^{k_0}_\mathcal{S}}). 
    \end{eqnarray}
    To obtain the maximum value $\mathcal{F}_i(\rho)$, one further measures qubits in $\mathcal{S}\backslash\{1,i\}$ in the chosen optimal basis starting from each $\ket{\Psi^{k_0}_\mathcal{S}}$ in the ensemble $\{p_{k_0},\ket{\Psi_{\mathcal{S}}^{k_0}}\}$, such that 
    \begin{eqnarray}
        \mathcal{F}_i(\rho)=\sum_{k_0} p_{k_0} f^{k_0}_i,
    \end{eqnarray}
    with 
    \begin{eqnarray}
        f^{k_0}_i=\underset{ \mathcal{M}_{\mathcal{S}\backslash\{1,i\}}}{\text{max}}\sum_k q_{(k_0,k)} E(\ket{\Psi^{(k_0,k)}_{1i}}). 
    \end{eqnarray}    
    Here, $\{q_{k_0,k},\ket{\Psi^{(k_0,k)}_{1i}}\}$ is the post-measured ensemble on the qubit-pair $\{1,i\}$, obtained by measuring qubits belonging to $\mathcal{S}\backslash\{1,i\}$ in the state $\ket{\Psi^{k_0}_\mathcal{S}}$. Using monotonicity of entanglement, 
    \begin{eqnarray}
        f^{k_0}_i \leq \min\left\{E_{1:23\cdots N}(\ket{\Psi^{k_0}_\mathcal{S}}),E_{2:13\cdots N}(\ket{\Psi^{k_0}_\mathcal{S}})\right\}.
    \end{eqnarray}
    Further, noticing that each $\ket{\Psi^{k_0}_{\mathcal{S}}}$ is permutation-symmetric,  $f^{k_0}_i \leq E_{1:23\cdots N}(\ket{\Psi^{k_0}_\mathcal{S}})$, leading to $\mathcal{F}_i(\rho)\leq \mathcal{E}^\prime_\mathcal{S}(\rho)$. Since $\mathcal{E}^\prime_\mathcal{S}(\rho)\leq \mathcal{E}_\mathcal{S}(\rho)$ by definition, $\mathcal{F}_i(\rho)\leq \mathcal{E}_\mathcal{S}(\rho)$, thereby proving Proposition 2 in the same line as Proposition 1.  
\end{proof}

\begin{figure}
    \centering
    \includegraphics[width=0.9\linewidth]{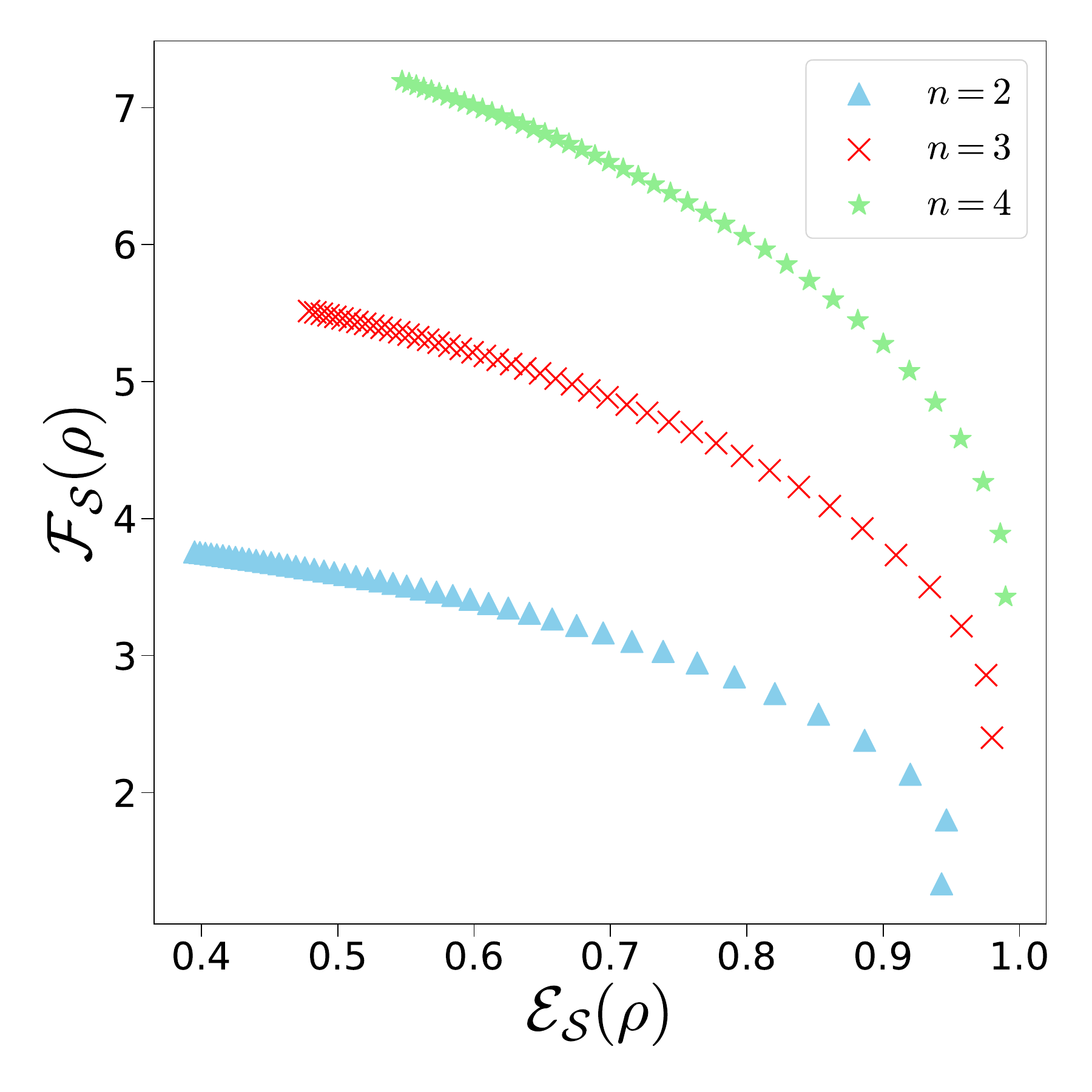}
    \caption{\textbf{$\mathcal{F}_\mathcal{S}(\rho)$ (ordinate) as a function of $\mathcal{E}_\mathcal{S}(\rho)$ (abscissa) for Dicke states \(|D_m\rangle\) (Eq.~(\ref{eq:dicke_computational_basis}))
    of different system-sizes $M\leq50$.} The BLE,  $\mathcal{E}_\mathcal{S}(\rho)$, and the total RLE, $\mathcal{F}_\mathcal{S}(\rho)$, can be obtained from  Eqs.~(\ref{eq:s0_non_empty_dicke_E}) and (\ref{eq:eq:s0_non_empty_dicke_F}) respectively. 
    Here excitations $n=2,3$ and $4$ are taken, and \(\mathcal{S}_0\) contains one qubit. All quantities  are dimensionless}
    \label{fig:s0_non_empty_dicke}
\end{figure}

\noindent For $M$-qubit W states i.e., the Dicke states of single excitations,  $\mathcal{F}_i(\rho)=2/M$ and $\mathcal{E}_\mathcal{S}(\rho)=2\sqrt{N-1}/M$ leading to $\mathcal{F}_\mathcal{S}(\rho)=\sqrt{N-1}\mathcal{E}_\mathcal{S}(\rho)$. For $M$-qubit Dicke states with arbitrary $n$\footnote{This expression can be derived using results from~\cite{Krishnan2023}, and working it out for systems with small sizes, and subsequently generalizing the results for arbitrary $N$.},
\begin{eqnarray}
\label{eq:s0_non_empty_dicke_E}
    \mathcal{E}_\mathcal{S}(\rho)=\frac{2}{N\genfrac(){0pt}{1}{M}{n}}\sum_{l=\text{max}(0,n-M+N)}^{\text{min}(n,N)} \genfrac(){0pt}{1}{N}{l}\genfrac(){0pt}{1}{M-N}{n-l} \sqrt{l(N-l)},
\end{eqnarray}
and 
\begin{eqnarray}
    \mathcal{F}_i(\rho)=\frac{2n(M-n)}{M(M-1)},
\end{eqnarray}
leading to
\begin{eqnarray}
\label{eq:eq:s0_non_empty_dicke_F}
    \mathcal{F}_\mathcal{S}(\rho)=\frac{2n(M-n)}{M(M-1)}(N-1),
\end{eqnarray} 
and subsequently $\mathcal{E}_\mathcal{S}(\rho)<\mathcal{F}_\mathcal{S}(\rho)<(N-1)\mathcal{E}_\mathcal{S}(\rho)$. For the case of $\mathcal{S}_0$ constituted of only one qubit, the dependence of $\mathcal{F}_\mathcal{S}(\rho)$ on  $\mathcal{E}_\mathcal{S}(\rho)$ for Dicke states with different $n$ is demonstrated in Fig.~\ref{fig:s0_non_empty_dicke}. For the $M$-qubit gGHZ states (Eq.~(\ref{eq:gGHZ_state})),  $\mathcal{E}_\mathcal{S}(\rho)=\mathcal{F}_i(\rho)=2|c_0|\sqrt{1-|c_0|^2}$, leading to the saturation of the upper bound, i.e., $\mathcal{F}_\mathcal{S}(\rho)=(N-1)\mathcal{E}_\mathcal{S}(\rho)$, while $\mathcal{E}_\mathcal{S}(\rho)<\mathcal{F}_\mathcal{S}(\rho)$. For the specific case of the GHZ state with $\mathcal{E}_\mathcal{S}(\rho)=1$, $\mathcal{F}_\mathcal{S}(\rho)=(N-1)$. For the more general family of single parameter states defined in Eq.~(\ref{eq:symmetric_state_magnetization_pair}), our numerical analysis suggests   $\mathcal{E}_{\mathcal{S}}(\rho)<\mathcal{F}_{\mathcal{S}}(\rho)\leq (N-1)\mathcal{E}_{\mathcal{S}}(\rho)$.

\subsection{States from a specific magnetization sector}
\label{subsec:magnetization_sector_1}

We start with $\ket{\psi_m}$ for $n=1$ and arbitrary $N$, also known as the  $N$-qubit generalized W (gW) state~\cite{dur2000,sende2003}, given in the computational basis as 
\begin{eqnarray}
\ket{\text{gW}}_\mathcal{S} &=& \sum_{i=1}^N c_i\ket{0}^{\otimes (i-1)}\ket{1}_i\ket{0}^{\otimes (N-i)},
\label{eq:n_qubit_gw}
\end{eqnarray}
where $c_i\in\mathbb{C}$ $\forall i\in\{1,2,\cdots,N\}$, satisfying the normalization condition $\sum_{i=1}^N|c_i|^2=1$.  For these states, according to our numerical analysis, the optimization of each of the components of $\mathcal{F}_{\mathcal{S}}(\rho)$ takes place via Pauli measurements on each qubit when $n=1$ (c.f.~\cite{Krishnan2023}). The following Propositions 3 and 4 describe respectively the upper and the lower bounds of $\mathcal{F}_{\mathcal{S}}(\rho)$ in terms of $E(\rho)$ when $\mathcal{S}_0=\emptyset$.  

\propose{3} \emph{For all $N$-qubit gW states, $\mathcal{F}_{\mathcal{S}}(\rho)\leq \sqrt{N-1}E(\rho)$, where the bound is saturated by a family of gW states with}
\begin{eqnarray}
    |c_i|^2=(N-1)^{-1}\left[1-|c_1|^2\right];\;\forall\;i \in\{2,\cdots,N\}.
    \label{eq:A_gW_UB_condition}
\end{eqnarray}

\begin{proof}
For $N$-qubit gW states,  $E(\rho)=|c_1|\sqrt{1-|c_1|^2}$ and 
\begin{eqnarray}
    \mathcal{F}_\mathcal{S}(\rho)=|c_1|\sum_{i=2}^N|c_i|
    \label{eq:A_FS}
\end{eqnarray}
Noting that a constant $E(\rho)$ is decided by a constant $|c_1|$, maximizing $\mathcal{F}_{\mathcal{S}}(\rho)$ for a fixed $E(\rho)$ reduces to maximizing $\sum_{i=2}^{N}|c_i|$ with the normalization constraint, i.e., $\sum_{i=2}^{N}|c_i|^2=1-|c_1|^2$. Since $|c_i|\geq0$ $\forall i=2,\cdots,N$, maximizing $\sum_{i=2}^{N}|c_i|$ requires maximizing $|c_i|$ individually with the normalization constraint, which occurs when 
\begin{eqnarray}
    |c_i|=\left[\frac{1-|c_1|^2}{N-1}\right]^{\frac{1}{2}},\;\forall i\in\{2,3,\cdots,N\}. 
\end{eqnarray}
The maximum value of $\mathcal{F}_{\mathcal{S}}(\rho)$, therefore, is 
\begin{eqnarray}
    \mathcal{F}_{\mathcal{S}}(\rho)=|c_1|\sqrt{\left(1-|c_1|^2\right)(N-1)}, 
\end{eqnarray}
leading to the line $\mathcal{F}_{\mathcal{S}}(\rho)=\sqrt{N-1}E(\rho)$.
Hence the proof. 
\end{proof}

\begin{figure*}
    \centering
    \includegraphics[width=0.8\linewidth]{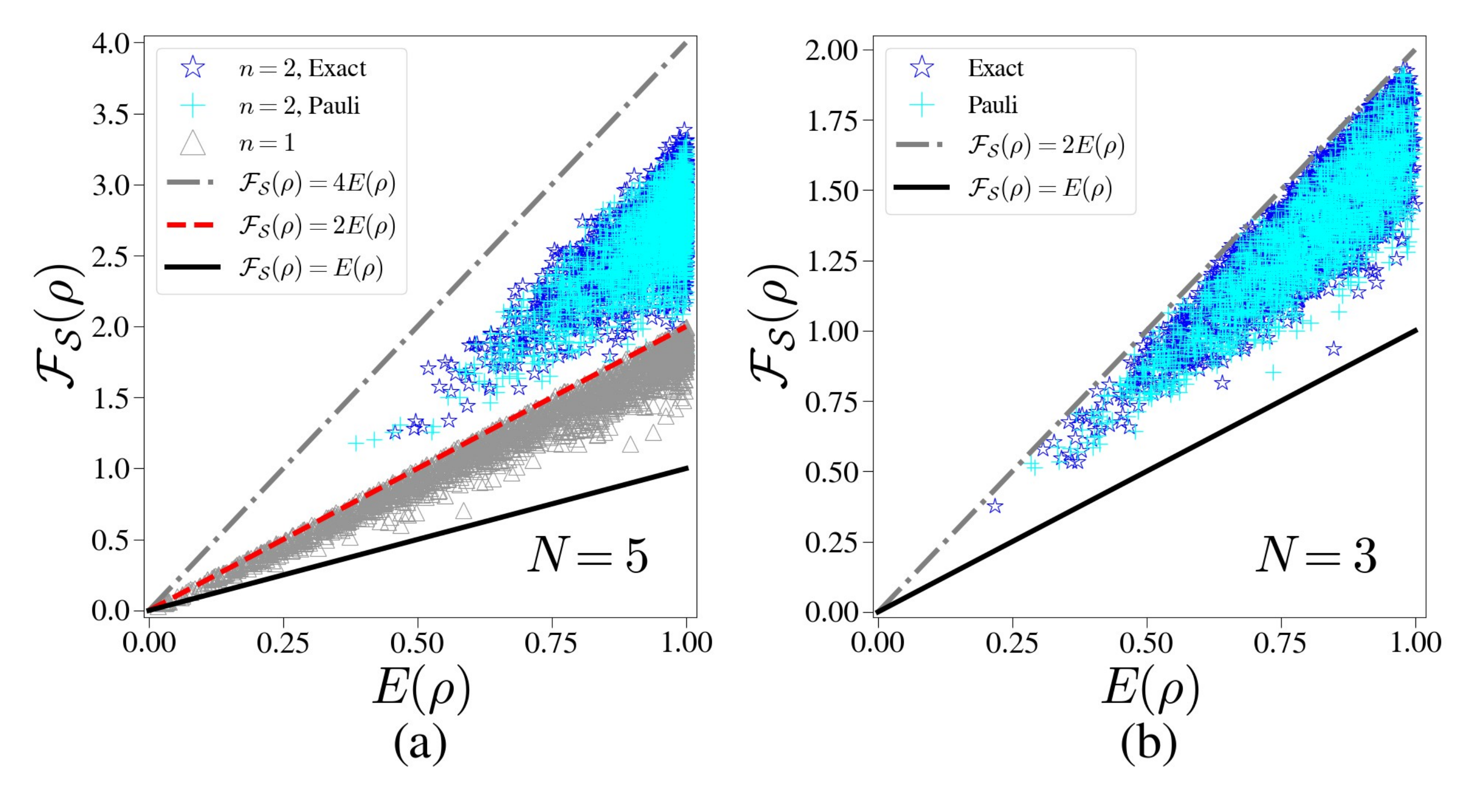}
    \caption{(a) \textbf{Scatter plot for 
    the states \(|\psi_m\rangle\) (Eq.~(\ref{eq:magnetization_sector_generic_state})) from specific magnetization sector  on the $\left(E(\rho),\mathcal{F}_{\mathcal{S}}(\rho)\right)$-plane with $\mathcal{S}_0=\emptyset$.} A set of randomly generated $5$-qubit gW states ($n=1$) lie within the lines $\mathcal{F}_{\mathcal{S}}(\rho)=2E(\rho)$ and $\mathcal{F}_{\mathcal{S}}(\rho)=E(\rho)$, as in Propositions 3 and 4 respectively. In contrast, $5$-qubit states of the form $\ket{\psi_m}$ with 
    two excitations ($n=2$) obeys the bound $\mathcal{F}_{\mathcal{S}}(\rho)\leq 4E(\rho)$, which is proposed in Proposition 1 for symmetric states. Samples of $10^6$ randomly generated $5$-qubit states  $\ket{\psi_m}$ for each of  $n=1$ and $n=2$ are used to generate the scatter plot.   (b) \textbf{Map of the three-qubit GHZ class states on the $\left(E(\rho),\mathcal{F}_{\mathcal{S}}(\rho)\right)$-plane with $\mathcal{S}_0=\emptyset$.} A set of $10^6$ Haar-uniformly generated three-qubit (GHZ class) states are bounded above and below by the lines $\mathcal{F}_{\mathcal{S}}(\rho)=2E(\rho)$ and $\mathcal{F}_{\mathcal{S}}(\rho)=E(\rho)$ respectively. Here \(\mathcal{S}_0 = \emptyset\). All quantities plotted are dimensionless in both (a) and (b).}
    \label{fig:s0_empty_non_symmetric}
\end{figure*}

\propose{4} \emph{For all $N$-qubit gW states,  $\mathcal{F}_\mathcal{S}(\rho)\geq E(\rho)$, with the family of states saturating the lower bound being either fully separable, or  $(N-1)$-separable\footnote{An $N$-qubit state $\rho$ is $L$-separable ($2\leq L\leq N$) if it can be written as the product of at-least $L$ states $\varrho_i$, $i=1,2,\cdots,L$,  corresponding to $L$ \emph{mutually exclusive} partitions of the $N$-qubit system such that the union of all $L$ partitions constitute the $N$-qubit system.}}.

\begin{proof}
To prove this Proposition, we start from Eq.~(\ref{eq:A_FS}), and note that fixing $E(\rho)$ implies assuming $|c_1|$, and  subsequently $\sum_{i=2}^N|c_i|^2=1-|c_1|^2$ to be constants. Defining $C=\sum_{i=2}^N|c_i|$, note that minimizing $\mathcal{F}_{\mathcal{S}}(\rho)$ amounts to minimizing $C$, which leads to a fully separable $N$-qubit state for which $\mathcal{F}_{\mathcal{S}}(\rho)=E(\rho)=0$ (since $|c_i|\geq 0$, $C=0$ implies $|c_i|=0$ $\forall i=2,3,\cdots,N$). Now consider the case of $C>0$, which can occur if $K$ ($1\leq K\leq N-1$) of $\{|c_i|;i=2,3,\cdots,N\}$ are $>0$, with values satisfying the normalization condition. Consider the minimal case of $K=1$, where $|c_j|>0$, and $|c_{i\neq j}|=0$ $\forall i=2,3,\cdots,N$, which corresponds to the $(N-1)$-separable states
\begin{eqnarray}\label{eq:gW_LB_state}
    \ket{\text{gW}}=\left(c_1\ket{1}_1\ket{0}_i+c_j\ket{0}_1\ket{1}_j\right)\bigotimes_{\underset{i\neq j}{i=2}}^N\ket{0}_i, 
\end{eqnarray}
for which 
\begin{eqnarray}
    \mathcal{F}_{\mathcal{S}}(\rho)=|c_1||c_j|=|c_1|\sqrt{1-|c_1|^2}
    = E(\rho). 
\end{eqnarray}
Next, consider the case of $K=2$, where  $|c_j|, |c_k|>0$, and $|c_{i\neq j,k}|=0$ $\forall i=2,3,\cdots,N$, such that $\mathcal{F}_{\mathcal{S}}(\rho)=|c_1|(|c_j|+|c_k|)$, and $|c_j|^2+|c_k|^2=1-|c_1|^2$. Assuming that $\mathcal{F}_{\mathcal{S}}(\rho)\leq E(\rho)$ leads to $|c_j||c_k|\leq 0$, which is a contradiction, thereby proving $\mathcal{F}_{\mathcal{S}}(\rho)>E(\rho)$. The same line of arguments holds for all cases of $2\leq K\leq N-1$. Hence the proof. 
\end{proof}
\noindent In Fig.~\ref{fig:s0_empty_non_symmetric}(a), these lower and upper bounds are demonstrated for randomly generated five-qubit gW states, where $c_i=c_i^\prime+\text{i}c_i^{\prime\prime}$, and the real parameters $c_i^\prime$ and $c_i^{\prime\prime}$ are chosen from a Gaussian distribution of zero mean and standard deviation $1$. In contrast, for $n>1$ with fixed $N$, only the upper bound  $\mathcal{F}_{\mathcal{S}}(\rho)\leq (N-1)E(\rho)$ is found to be obeyed, as shown in Fig.~\ref{fig:s0_empty_non_symmetric}(a) by randomly generated $5$-qubit pure states of the form $\ket{\psi_m}$ with $n=2$.

For $\mathcal{S}_0\neq\emptyset$, our numerical analysis reveals that the optimal measurement for maximizing both $\mathcal{E}_\mathcal{S}(\rho)$ and $\mathcal{F}_i(\rho)$ $\forall i$ is $\sigma^z$. In the case of $n=1$, the gW state of $M$-qubits can be written as
\begin{eqnarray}
    \ket{\text{gW}}=\ket{\text{gW}}_{\mathcal{S}_0} \otimes \ket{0}^{\otimes N} + \ket{0}^{\otimes N_0} \otimes\ket{\text{gW}}_\mathcal{S}, 
\end{eqnarray}
where $\ket{\text{gW}}_\mathcal{S}$ is defined in Eq.~(\ref{eq:n_qubit_gw}) with the normalization condition $\sum_{i\in \mathcal{S}_0}|c_i|^2+\sum_{i\in \mathcal{S}}|c_i|^2=1$. The upper and lower bounds of $\mathcal{F}_\mathcal{S}(\rho)$ in terms of $\mathcal{E}_\mathcal{S}(\rho)$ are described respectively by the following Propositions 5 and 6.

\propose{5} \emph{For all $M$-qubit gW states, $\mathcal{F}_{\mathcal{S}}(\rho)\leq \sqrt{N-1}\mathcal{E}_\mathcal{S}(\rho)$, where the bound is saturated by a family of gW states with}
\begin{eqnarray}
    |c_i|^2=(N-1)^{-1}\left[1-|c_1|^2-\sum_{j\in \mathcal{S}_0}|c_j|^2\right],
    \label{eq:B_gW_UB_condition}
\end{eqnarray}
\emph{where $\forall~i \in\{2,\cdots,N\}$.}
\begin{proof}
    For the $M$-qubit gW states~\cite{Krishnan2023}, 
    \begin{eqnarray}
        \mathcal{E}_\mathcal{S}(\rho)=|c_1|\sqrt{1-|c_1|^2-\sum_{j\in \mathcal{S}_0}|c_j|^2},
    \end{eqnarray}
    and $\mathcal{F}_\mathcal{S}(\rho)$ is given by Eq.~(\ref{eq:A_FS}). One can make $\mathcal{E}_\mathcal{S}(\rho)$ constant by fixing $|c_1|$ and $\sum_{j \in \mathcal{S}_0}|c_j|^2$, which,  due to normalization, results in 
    \begin{eqnarray}\label{eq:B_FS}
        \sum_{i=2}^N|c_i|^2 = 1-|c_1|^2-\sum_{j\in \mathcal{S}_0}|c_j|^2 . 
    \end{eqnarray}
    Maximization of Eq.~(\ref{eq:A_FS}) reduces to the maximization of $\sum_{i=2}^N |c_i|$  subjected to the constraints~(\ref{eq:B_FS}) and $|c_i|\geq 0$ $\forall~i=2,\cdots,N$. Consequently, individual maximization of $|c_i|$s results in 
    \begin{eqnarray}
        |c_i|=\left[\frac{1-|c_1|^2-\sum_{j\in \mathcal{S}_0}|c_j|^2}{N-1}\right]^\frac{1}{2},
    \end{eqnarray}
    and the maximum value of $\mathcal{F}_{\mathcal{S}}(\rho)$ can be subsequently computed as  
    \begin{eqnarray}
        \mathcal{F}_{\mathcal{S}}(\rho)=|c_1|\sqrt{\left(1-|c_1|^2-\sum_{j \in \mathcal{S}_0}|c_j|^2\right)(N-1)},
    \end{eqnarray}
    which leads to the upper bound $\mathcal{F}_{\mathcal{S}}(\rho)=\sqrt{N-1}\mathcal{E}_\mathcal{S}(\rho)$.
\end{proof}

\begin{figure*}
    \centering
    \includegraphics[width=0.8\linewidth]{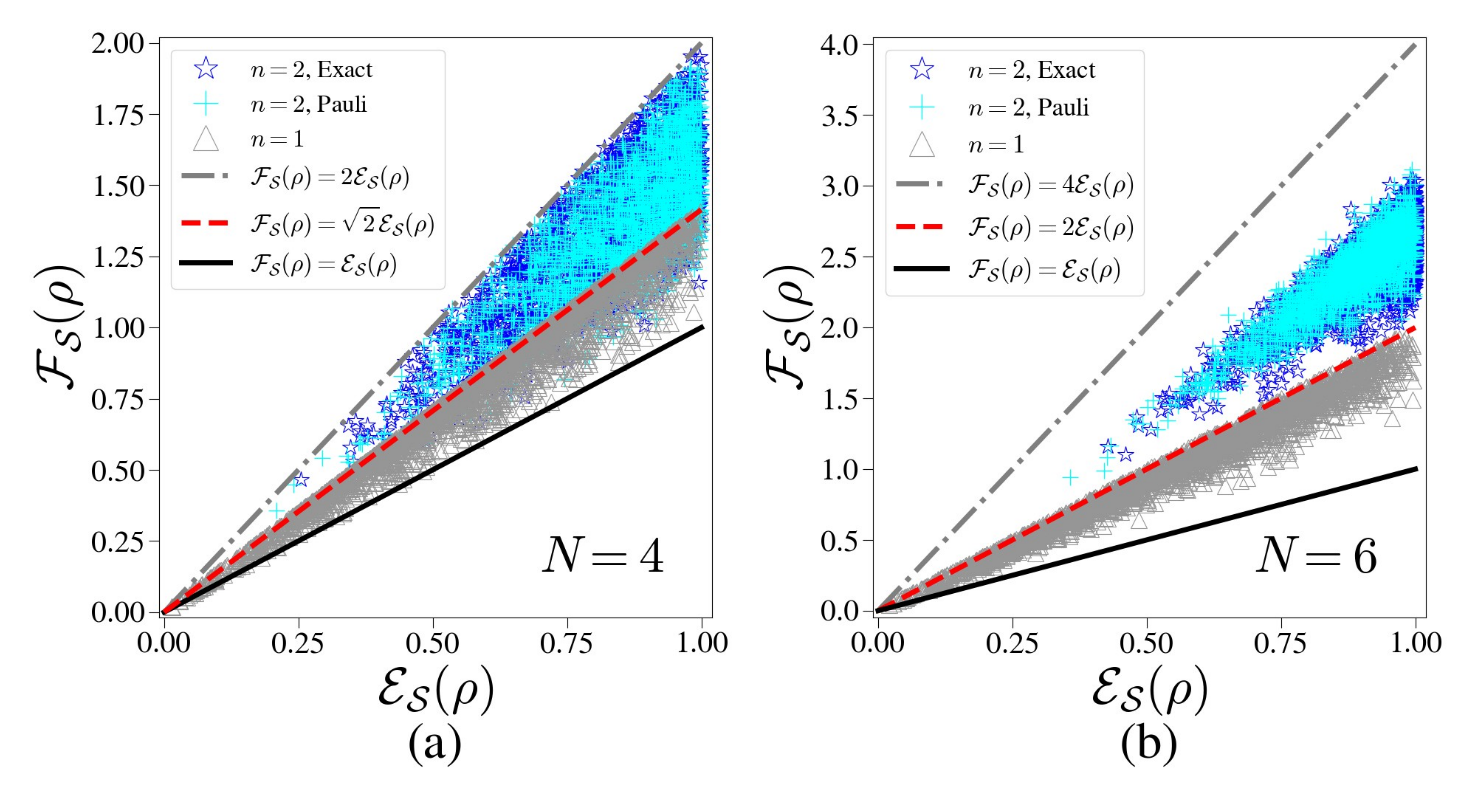}
    \caption{\textbf{Scatter plot for 
    the states \(|\psi_m\rangle\) (Eq.~(\ref{eq:magnetization_sector_generic_state})) from specific magnetization sector  on the $\left(E(\rho),\mathcal{F}_{\mathcal{S}}(\rho)\right)$-plane with $\mathcal{S}_0\neq\emptyset$.} (a) A set of $10^6$  randomly generated  four-qubit gW states ($n=1$) lie within the lines $\mathcal{F}_{\mathcal{S}}(\rho)=\sqrt{2}\mathcal{E}_\mathcal{S}(\rho)$ and $\mathcal{F}_{\mathcal{S}}(\rho)=\mathcal{E}_\mathcal{S}(\rho)$, obeying Propositions 5 and 6 respectively. However, four-qubit states $\ket{\psi_m}$ with $n=2$ obeys the bound $\mathcal{F}_{\mathcal{S}}(\rho)\leq 2\mathcal{E}_\mathcal{S}(\rho)$. (b) Map of six-qubit states $\ket{\psi_m}$ with $n=1$ and $2$. The six-qubit gW state ($n=1$) falls between the respective bounds described in Propositions 5 and 6, while the states $\ket{\psi_m}$ with $n=2$ are obeying the bound $\mathcal{F}_{\mathcal{S}}(\rho)\leq 4\mathcal{E}_\mathcal{S}(\rho)$. In both (a) and (b), $\mathcal{S}_0$ hosts a single qubit. Optimizations over measurement basis is performed by maximizing over all parameters (exact) and by restricting the measurements only to  Pauli  basis (marked as Pauli), both of which  are shown to have similar trends. All quantities plotted are dimensionless.}
\label{fig:s0_nonempty_non_symmetric}
\end{figure*}

\propose{6} \emph{For arbitrary $M$-qubit gW states,  $\mathcal{F}_\mathcal{S}(\rho)\geq \mathcal{E}_\mathcal{S}(\rho)$, with the family of states saturating the lower bound being either $N$ separable, or  $(N-1)$-separable}.
\begin{proof}
    Fixing $|c_1|$ and $\sum_{j\in \mathcal{S}_0}|c_j|^2$ fixes $\mathcal{E}_\mathcal{S}(\rho)$,  leading to $\sum_{i=2}^N|c_i|^2=1-|c_1|^2-\sum_{j\in \mathcal{S}_0}|c_j|^2$. Let us now define $C=\sum_{i=2}^N|c_i|$, which has to be minimized for minimizing $\mathcal{F}_{\mathcal{S}}(\rho)$. When $C=0$, $|c_i|\geq0$ implies $|c_i|=0$  $\forall~i=2,\cdots ,N$. This results in an $N$ separable state of the form
    \begin{eqnarray}  
       \ket{\text{gW}}_0=\left(\ket{\text{gW}}_\mathcal{S}\otimes \ket{0}_1+\ket{0}^{\otimes N_0}\otimes c_1 \ket{1}_1\right)\otimes \ket{0}^{\otimes N-1},\nonumber \\
    \end{eqnarray}
    for which $c_1$ is determined by the normalization condition, and $\mathcal{E}_\mathcal{S}(\rho)=\mathcal{F}_\mathcal{S}(\rho)=0$. On the other hand,  $C>0$  when $K$ $(1\leq K\leq (N-1))$ of the $\{|c_i|; i=2,3\cdots ,N \}$ are $>0$, with their values satisfying the normalization condition. Considering the minimal case of $K=1$, where $|c_j|>0$ and $|c_{j\neq i|}=0$ $\forall~i=2,3\cdots ,N$, which corresponds to the ($N-1$) separable states
    \begin{eqnarray}
        \ket{\text{gW}}&=&\Big[\ket{\text{gW}}_{\mathcal{S}_0}\ket{0}_1\ket{0}_j +\ket{0}^{\otimes N_0}\Big(c_1\ket{1}_1\ket{0}_j \nonumber \\ && + c_j\ket{0}_1\ket{1}_j\Big)\Big] \bigotimes_{\underset{i\neq j}{i=2}}^N\ket{0}_i,
    \end{eqnarray}
    with $c_1$ and $c_j$ constrained by the normalization,  and
    \begin{eqnarray}
        \mathcal{F}_{\mathcal{S}}(\rho)&=&|c_1||c_j|=|c_1|\sqrt{1-|c_1|^2-\sum_{j\in \mathcal{S}_0}|c_j|^2}\nonumber\\&=& \mathcal{E}_\mathcal{S}(\rho). 
    \end{eqnarray}
    For $2\leq K \leq N-1$, the corresponding states can be shown to lie above the line $\mathcal{F}_{\mathcal{S}}(\rho)=\mathcal{E}_{\mathcal{S}}(\rho)$ following arguments similar to that from the proof of Proposition 4,  showing that $\mathcal{F}_{\mathcal{S}}(\rho)\geq\mathcal{E}_{\mathcal{S}}(\rho)$. Hence, the proof.
\end{proof}
Although the above propositions are presented for gW states (i.e., with a single excitation), we numerically observe that bounds different than the gW states emerge  with the increase of the number of excitations in the states of the form $\ket{\psi_m}$, in both cases of $\mathcal{S}_0=\emptyset$ and $\mathcal{S}_0\neq\emptyset$, as depicted for $n=2$ in Figs.~\ref{fig:s0_empty_non_symmetric}(a) and~\ref{fig:s0_nonempty_non_symmetric} (a), and (b). Note that the bounds for $n>1$ resembles the bound obtained for permutation-symmetric states.

\subsection{Arbitrary multi-qubit pure states} 
\label{subsec:arbitrary_states}

We now consider arbitrary $N$-qubit pure states, and first turn towards the specific case of $N=3$. The full set of three-qubit pure states can be divided into two mutually disjoint classes of states, namely, the GHZ and the W classes~\cite{dur2000}. A state from the former class can be written as $\ket{\Phi_{\text{GHZ}}}=\sum_{l=0}^7c_l\ket{b_l}$,
with $c_l$ complex and $\sum_{l=0}^{7}|c_l|^2=1$, while a generic W class state can be written as $\ket{\Phi_{\text{W}}} =c_0\ket{b_0^3}+c_1\ket{b_4^1}+c_2\ket{b_2^1}+c_3\ket{b_1^1}$, 
where $c_i$'s are complex with $\sum_{i=0}^3|c_i|^2=1$, and the basis $\{\ket{b_l^m}\}$ is sorted according to the decreasing value of $m$ from $+3$ to $-3$. In the computational basis, $\ket{b_0^3}=\ket{000}$, $\ket{b_1^1}=\ket{001}$, $\ket{b_2^1}=\ket{010}$, and $\ket{b_4^1}=\ket{100}$. Note that the three-qubit gGHZ and gW states are special cases of the three-qubit GHZ and W classes, respectively. The optimizations of the components of $\mathcal{F}_{\mathcal{S}}(\rho)$ take place corresponding to Pauli measurements on all qubits in the case of the three-qubit W class states also (cf.~\cite{Krishnan2023}), and  the following Propositions hold. 

\propose{7} \emph{For three-qubit W class states, $\mathcal{F}_{\mathcal{S}}(\rho)\leq \sqrt{2}E(\rho)$, with the family of states saturating this upper bound given by}  
\begin{eqnarray}
    |c_2|^2=|c_3|^2=\frac{1-|c_0|^2-|c_1|^2}{2}.
\end{eqnarray}

\propose{8} \emph{For all three-qubit W class states,  $\mathcal{F}_\mathcal{S}(\rho)\geq E(\rho)$, with the family of states saturating the lower bound being either biseparable, or fully separable.}

\noindent Noting that $E(\rho)=|c_1|\sqrt{|c_2|^2+|c_3|^2}$ and $\mathcal{F}_\mathcal{S}(\rho)=|c_1|(|c_2|+|c_3|)$~\cite{Krishnan2023} for three-qubit W class states, the proofs of Propositions 7 and 8  follow the same line of arguments as that of Propositions 3 and 4 respectively, where in the present case, one needs to fix both $|c_0|$ and $|c_1|$ to fix $E(\rho)$. In contrast to the W class states, maximization  of different components of $\mathcal{F}_\mathcal{S}(\rho)$ takes place outside the realm of Pauli basis for the GHZ class state, and hence analytical computation of $\mathcal{F}_\mathcal{S}(\rho)$ becomes cumbersome. However, our extensive numerical analysis suggests that the lower and upper bounds, given respectively by  $\mathcal{F}_\mathcal{S}(\rho)\geq E(\rho)$ and $\mathcal{F}_\mathcal{S}(\rho)\leq 2E(\rho)$, are obeyed by the arbitrary three-qubit GHZ class states, as demonstrated in Fig.~\ref{fig:s0_empty_non_symmetric}(b).

\begin{figure*}
    \centering
    \includegraphics[width=\linewidth]{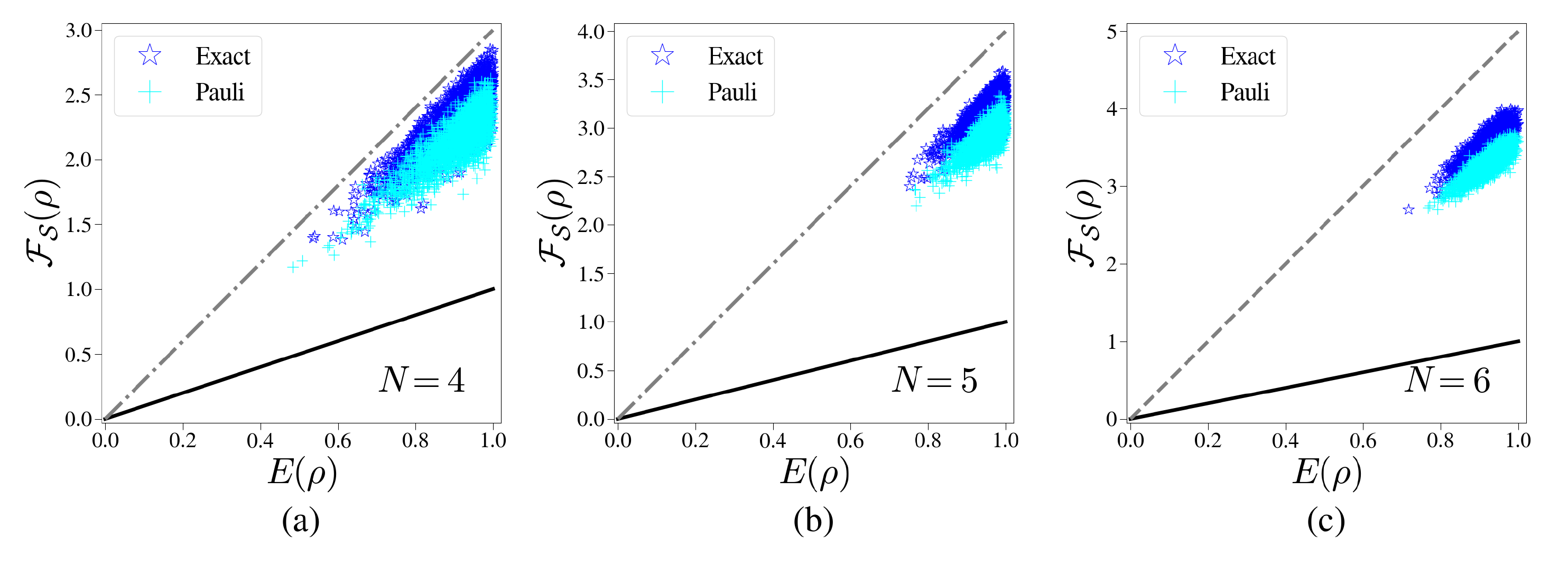}
    \caption{\textbf{Scatter plot of Haar uniformly generated multi-qubit states on the $\left(E(\rho),\mathcal{F}_\mathcal{S}(\rho)\right)$-plane.} A set of Haar uniformly generated $10^6$ states for each of (a) $N=4$, (b) $N=5$ and (c) $N=6$ is used to generate the plot.  Here \(\mathcal{S}_0=\emptyset\), while
    all other specifications are same as in Fig. \ref{fig:s0_nonempty_non_symmetric}. All quantities plotted are dimensionless.}
    \label{fig:s0_empty_haar_uniform}
\end{figure*}

Going beyond three-qubit states, the upper and lower bounds obtained for the GHZ class states in terms of $\mathcal{F}_{\mathcal{S}}(\rho)$ and $E(\rho)$ turn out to be weak when the system-size grows. In particular, by generating four-, five-, and six-qubit states Haar uniformly, we compute $\mathcal{F}_\mathcal{S}(\rho)$ by optimizing over all measurements and by restricting to Pauli measurements, and $E_\mathcal{S}(\rho)$. Clearly, $E_\mathcal{S}(\rho)\leq \mathcal{F}_\mathcal{S}(\rho)\leq 2E_\mathcal{S}(\rho)$ (see Fig.~\ref{fig:s0_empty_haar_uniform} and Proposition 1). It is worthwhile to point out that the bounds derived from the permutation-symmetric states hold also for Haar-uniformly generated arbitrary multi-qubit pure states. In contrast, the bounds derived for the arbitrary superposition of states from a specific magnetization sector are completely distinct from the former, highlighting their fundamentally different entanglement properties.

\section{States under local dephasing noise}
\label{sec:local_dephasing}

The entire analysis performed in Sec.~\ref{sec:pure_states} provides a relation between the total RLE, the BLE, and the BE for generic as well as paradigmatic classes of pure states having specific features. However, preparing pure states is an ideal situation. Hence, it is important to determine how these bounds get disturbed when   
each qubit of the multi-qubit system is sent through independent phase-flip channels~\cite{nielsen2010,holevo2012}. These channels can be either Markovian~\cite{yu2009}, or non-Markovian~\cite{Daffer2004,Shrikant2018,Gupta2020} in nature. Using operator-sum representation~\cite{nielsen2010}, the state of the system under noise is given by $ \rho = \sum_{\alpha}K_\alpha \rho_0 K_\alpha^\dagger,$
%
where $\rho_0$ is the multi-qubit pure state under consideration, and $\{K_\alpha\}$ are the Kraus operators, satisfying $\sum_{\alpha}K_\alpha^\dagger K_\alpha = I$, $I$ being the identity operator in the Hilbert space of the multi-qubit system. Here,  $K_\alpha=\sqrt{p_\alpha}K^\prime_\alpha$ with $K^\prime_\alpha =\bigotimes_{i\in\mathcal{S}_0\cup\mathcal{S}} K^\prime_{\alpha_i};\;p_\alpha = \prod_{i\in \mathcal{S}_0\cup\mathcal{S}} p_{\alpha_i}$,
where $\alpha$ is interpreted as a multi-index, $\sum_{\alpha_i}p_{\alpha_i}=1$.  In the case of the phase-flip noise for which $\alpha_i\in\{0,1\}$, the form of $K^\prime_{\alpha_i}$ is given by $K^\prime_{\alpha_i=1} = \sigma^z_i,\; K^\prime_{\alpha_i=0} = I_i$~\cite{nielsen2010},
with $\{p_{\alpha_i=0} = 1-\frac{q}{2},\;p_{\alpha_i=1} = \frac{q}{2}\}$
%
in the Markovian case~\cite{yu2009}, and 
\begin{eqnarray}\label{eq:noise_prob}
p_{\alpha_i=0} &=& \left(1-\frac{q}{2}\right)\left(1-\frac{\eta q}{2}\right),\nonumber\\
p_{\alpha_i=1} &=& \left[1+\eta\left(1-\frac{q}{2}\right)\right]\frac{q}{2}
\end{eqnarray}
in the non-Markovian case~\cite{Daffer2004,Shrikant2018,Gupta2020},  with the \emph{noise strength} denoted by $q$ ($0\leq q\leq 1$), and $\eta$ being the non-Markovianity parameter ($0\leq \eta\leq 1$) (with $\eta=0$ representing Markovian noise).

\subsection{Noisy permutation-symmetric states}
\label{subsec:noisy_permutation_symmetric}

\begin{figure*}
    \centering
    \includegraphics[width=\linewidth]{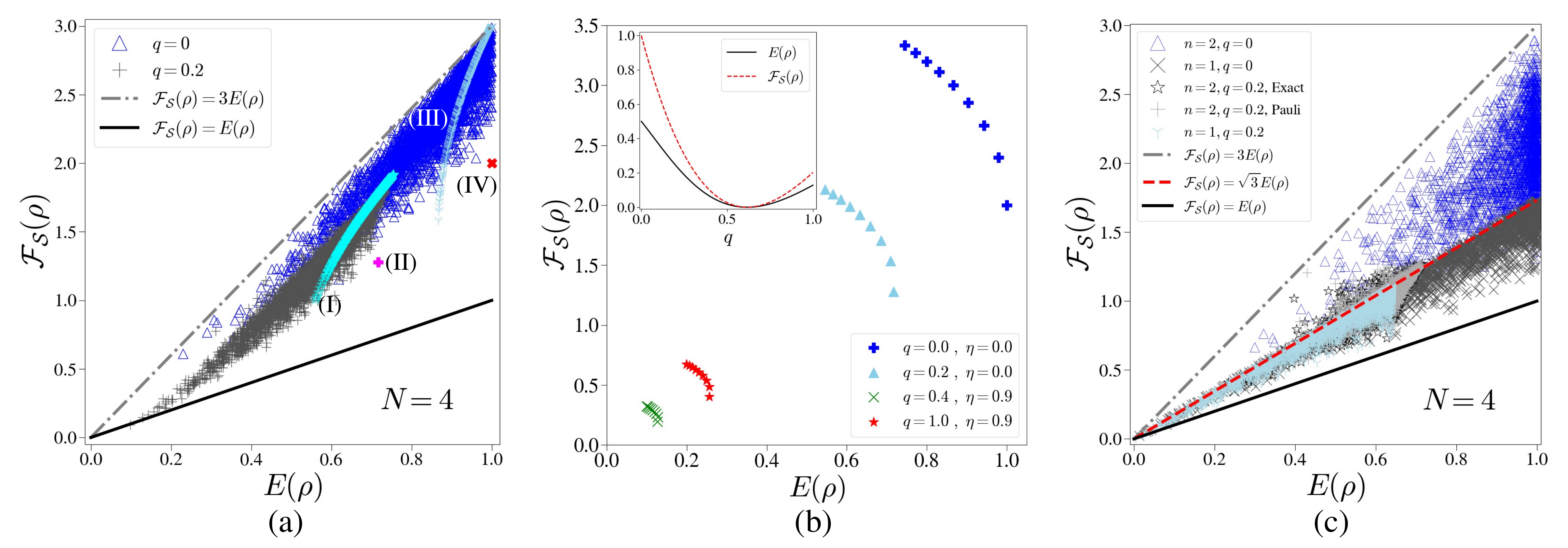}
    \caption{\textbf{Distribution of mixed states on the $\left(E(\rho),\mathcal{F}_\mathcal{S}(\rho)\right)$-plane with $\mathcal{S}_0=\emptyset$.} (a) Scatter plot of $10^6$ symmetric mixed states obtained by applying Markovian phase flip noise of strength $q=0.2$ to all qubits of a $4$-qubit state $\ket{\Psi}$ (Eq.~(\ref{eq:ghz_type})) (represented by crosses) on the background of the corresponding noiseless data (represented by hollow triangles). The dash dotted and continuous lines represent $\mathcal{F}_\mathcal{S}(\rho)=3E(\rho)$ and  $\mathcal{F}_\mathcal{S}(\rho)=E(\rho)$ respectively. The distribution of the single parameter state $\ket{\Psi_{(+2,-2)}}$ (Eq.~(\ref{eq:symmetric_state_magnetization_pair})) is labeled by (III), while (I) denotes its noisy counterpart. The points  (IV) and (II) respectively represent the state $\ket{D_0}$  (Eq.~(\ref{eq:dicke_computational_basis})) and the corresponding mixed state. (b) Scatter plot of Dicke states with different system-sizes $N\leq12$ but same number of excitations $n=2$, sent through non-Markovian local phase-flip channels. The non-Markovianity parameter is set at $\eta=0$ (Markovian) and $\eta=0.9$ (highly non-Markovian), while the noise strength is set at $q=0,0.2,0.4,1$. The revival of entanglement under the non-Markovian phase-flip noise can be observed from the position of the data corresponding to $q=1.0$ above the same for $q=0.4$, while it has been explicitly showed in the inset via the variations of  $E(\rho)$ and $\mathcal{F}_\mathcal{S}(\rho)$ with $q$ at $\eta=0.9$. At $q\approx0.62$, both $E(\rho)$ and $\mathcal{F}_\mathcal{S}(\rho)$ vanish. (c) Map of the states $\ket{\psi_m}$ from a specific magnetization sector (Eq.~(\ref{eq:magnetization_sector_generic_state})) under local phase-flip noise. A set of $10^6$  states $\ket{\psi_m}$ with $n=1$  and $n=2$ with $N=4$ are used to generate the data, keeping $q=0.2$ and $\eta=0$. The dash-dotted line, $\mathcal{F}_\mathcal{S}(\rho)=3E(\rho)$ and the dashed line, $\mathcal{F}_\mathcal{S}(\rho)=\sqrt{3}E(\rho)$ corresponds to the bounds defined in Propositions 1 and 3, whereas the solid line follows from Proposition 4. All quantities plotted are dimensionless.}
    \label{fig:noise_results}
\end{figure*}

For $\mathcal{S}_0=\emptyset$, all $N$-qubit  mixed states obtained from the permutation-symmetric pure states subjected to local uncorrelated phase-flip noise obey Proposition 1, since even after the action of local noise of the same strength on all qubits, the state remains permutation-symmetric. For the gGHZ states (Eq.~(\ref{eq:gGHZ_state})), under the action of Markovian and non-Markovian phase-flip noise of strength $q$, the entanglement of the noisy state  can be expressed in terms of the entanglement of the initial pure state~\cite{Krishnan2023},   and the upper bound put forward in Proposition 1 saturates. For the states of the form in Eq.~(\ref{eq:symmetric_state_magnetization_pair}), our numerical investigation suggests  that under the action of Markovian as well as non-Markovian phase flip noise, $E(\rho)<\mathcal{F}_\mathcal{S}(\rho)< (N-1)E(\rho)$. However, the most general form of permutation-symmetric states, as given in Eq.~(\ref{eq:ghz_type}), when sent through Markovian or non-Markovian phase-flip channels, are observed to violate $E(\rho)<\mathcal{F}_\mathcal{S}(\rho)$ for $0.0687\%$ ($q=0.2$ and $\eta=0$), i.e., for a small fraction of total states, $E(\rho)>\mathcal{F}_\mathcal{S}(\rho)$ (see  Fig.~\ref{fig:noise_results}(a)).

For $N$-qubit W state (Eq.~(\ref{eq:dicke_computational_basis}) with $n=1$) under non-Markovian phase-flip channels,  $E(\rho)=2(1-f(q,\eta))^2\sqrt{N-1}/N$ and $\mathcal{F}_i(\rho)=2(1-f(q,\eta))^2/N$~\cite{Krishnan2023}, where $f(q,\eta)$ is given by 
\begin{eqnarray}\label{eq:non_mark_fn}
    f(q,\eta)=q\left[1+\eta\left(1-\frac{q}{2}\right)\right], 
\end{eqnarray}
leading to $E(\rho)=\sqrt{N-1}\mathcal{F}_\mathcal{S}(\rho)$, implying a result similar to that of the noiseless scenario and a validity of the upper bound proved in Proposition 1 (note that the case of the Markovian noise can be derived from it by fixing $\eta=0$). In contrast, in the case of Dicke states with arbitrary excitations, tackling the state under the effect of Markovian as well as non-Markovian noise becomes analytically challenging. However, we perform numerical analysis for smaller system-sizes for fixed excitations  and for both Markovian and non-Markovian cases to find the upper bound to be valid in these cases as well. The behaviour of $\mathcal{F}_\mathcal{S}(\rho)$ with respect to $E(\rho)$ for the noisy Dicke states with $n=2$ and $N\leq 12$ is shown in Fig.~\ref{fig:noise_results}(b).

When $\mathcal{S}_0\neq\emptyset$, the $M$-qubit permutation-symmetric mixed states, obtained by sending the states through Markovian and non-Markovian phase flip channels, again obey Proposition 2 following a similar proof. Further, the results corresponding to the $M$-qubit gGHZ states in the noisy scenario are similar to that of the $\mathcal{S}_0=\emptyset$ case, where a saturation of the upper bound proposed in Proposition 2 is observed. 
Although the superposition of $\ket{D_m}$ and $\ket{D_{-m}}$ again follow
$\mathcal{E}_\mathcal{S}(\rho)<\mathcal{F}_\mathcal{S}(\rho)<(N-1)\mathcal{E}_\mathcal{S}(\rho)$, for noisy Dicke states generated with $n=1$ and $N=4$, $\mathcal{E}_\mathcal{S}(\rho)>\mathcal{F}_\mathcal{S}(\rho)$ for $0.1875 \%$ ($q=0.2$ and $\eta=0$) of total states. 

\begin{figure*}
    \centering
    \includegraphics[width=\linewidth]{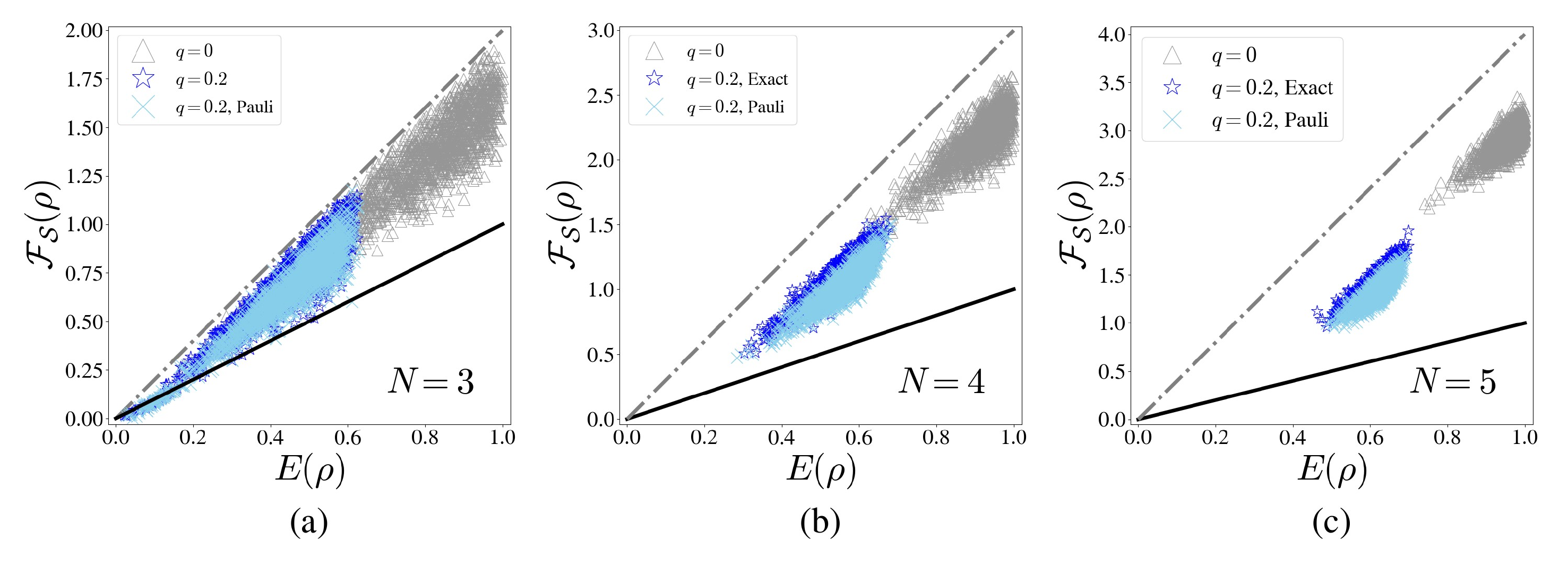}
    \caption{\textbf{Distribution of arbitrary multi-qubit states under local phase-flip channels.} Scatter plot of $10^6$ Haar-uniformly generated states with (a)  $N=3$ (GHZ class), (b) $N=4$ and (c) $N=5$, where all qubits are sent through a Markovian phase-flip channel with $q=0.2$. The lower bound is defined by (continuous lines) $\mathcal{F}_\mathcal{S}(\rho)=E(\rho)$ for all cases, while the upper bound (dash-dotted lines) is given by $\mathcal{F}_\mathcal{S}(\rho)=(N-1)E(\rho)$. The data includes LE obtained by both exact and Pauli optimizations, and the corresponding data for pure state is also included for comparison. All quantities plotted are dimensionless.}
    \label{fig:s0_empty_haar_uniform_noisy}
\end{figure*}

For $M$-qubit W states~(Eq.~(\ref{eq:dicke_computational_basis}) with $n=1$) under non-Markovian phase flip noise,  $\mathcal{E}_\mathcal{S}(\rho)=(1-f(q,\eta))^2\sqrt{N-1}/M$ and $\mathcal{F}_\mathcal{S}(\rho)=(1-f(q,\eta))^2(N-1)/M$~\cite{Krishnan2023}, leading to $\mathcal{F}_\mathcal{S}(\rho)=\sqrt{N-1}\mathcal{E}_\mathcal{S}(\rho)$, which is the same as the noiseless case, and remains unchanged in the Markovian limit ($\eta=0$) as well. Further, our numerical investigation on $M$-qubit Dicke states under Markovain and non-Markovian phase flip noise for smaller system-sizes, $M\leq 7$  confirms the validity of the upper bound proposed in Proposition 2. Also, our analysis reveals that the dependence of $\mathcal{F}_\mathcal{S}(\rho)$ on $\mathcal{E}_\mathcal{S}(\rho)$ for fixed $n$ is qualitatively similar to that of $\mathcal{F}_\mathcal{S}(\rho)$ on $E(\rho)$, as demonstrated for the latter case in Fig.~\ref{fig:noise_results}(b).

\subsection{Dephased states from specific magnetization sectors}
\label{subsec:noisy_specific_magnetization}

Starting with the gW state (Eq.~(\ref{eq:n_qubit_gw}) with $n=1$), variations of $E(\rho)$ and $\mathcal{F}_i(\rho)$ as functions of $q$ and $\eta$ can be derived under the action of Markovian and non-Markovian phase flip noise on all qubits~\cite{Krishnan2023}, from which the validity of the upper and lower bounds, given respectively by  $\mathcal{F}_\mathcal{S}(\rho)=\sqrt{N-1}E(\rho)$ and $\mathcal{F}_\mathcal{S}(\rho)=E(\rho)$, in Propositions 3 and 4 can be inferred. The family of states that lie on these lines can be obtained by sending the states in Eq.~(\ref{eq:A_gW_UB_condition}) (upper bound) and Eq.~(\ref{eq:gW_LB_state}) (lower bound) through Markovian or non-Markovian phase flip noise channel. Fig.~\ref{fig:noise_results}(c) demonstrates the distribution of states obtained by sending the states given in Eq.~(\ref{eq:magnetization_sector_generic_state}) through Markovian phase-flip channel  on  the $\left(E(\rho),\mathcal{F}_\mathcal{S}(\rho)\right)$-plane for $N=4$, where the states corresponding to $n=1$ can be seen to follow the upper bound described by $\mathcal{F}_\mathcal{S}(\rho)=\sqrt{N-1}E(\rho)$, whereas the states with $n>1$ obey the upper bound $\mathcal{F}_\mathcal{S}(\rho)=(N-1)E(\rho)$, as demonstrated for $n=2$. Qualitatively similar results are found when non-Markovianity is introduced in the noise. Further, bounds proposed in Propositions 5 and 6 are found to stand even in the case of $M$-qubit gW states sent through Markovian or non-Markovian phase-flip noise when $\mathcal{S}_0\neq\emptyset$ as well.

\subsection{Arbitrary states under phase-flip noise}
\label{subsec:noisy_arbitrary_states}

When $\mathcal{S}_0=\emptyset$ and the local phase-flip noise (with moderate noise strength) acts on the GHZ class states, we observe 
$E(\rho)>\mathcal{F}_\mathcal{S}(\rho)$ for the 0.0744 \%\ ($q=0.2$ and $\eta=0$) of the noisy states with noise strength \(q=0.2\) and \(\eta =0\) (see Fig.~\ref{fig:s0_empty_haar_uniform_noisy}(a) for a demonstration) although W class states obey the bound. However, the bounds remain intact when one considers Haar-uniformly~\cite{bengtsson2017} generated four- and five-qubit states. When $\mathcal{S}_0\neq \emptyset$, the distribution of Haar-uniformly~\cite{bengtsson2017} generated states under  Markovian as well as non-Markovian phase flip noises confirms the validity of the corresponding upper and lower bounds for pure states with an exception to $M=4$, which demonstrates $\mathcal{E}_\mathcal{S}(\rho)>\mathcal{F}_\mathcal{S}(\rho)$ for a negligible, yet non-zero percentage of total states sampled.

\section{Conclusion and outlook}
\label{sec:conclude}

Projective measurements serve a dual role in quantum systems: they can eliminate quantum correlations between measured and unmeasured parties, and yet can concentrate entanglement among the unmeasured ones. This characteristic makes them crucial tools in the development of quantum networks, where the goal is to establish robust entanglement among designated functional nodes. To facilitate this, we visualized a bipartition of the nodes of a quantum network into sets of useful and idle (inoperative) nodes, where concentrating entanglement over the useful nodes is possible by  performing suitable projective measurements on the idle ones. Considering one node to be the hub among the useful ones, which typically is the node with maximum connectivity in a network, we investigated the connection between the entanglement shared over the bipartition of the hub and the rest of the useful nodes obtained via measurement and partial trace over the idle nodes, referred to as the block entanglement, and the total localizable entanglement over all links converging on the hub, referred to as the total regionally localized entanglement. We analyzed and provided bounds on the total regionally localized entanglement in terms of the block entanglement computed using the measurement-based as well as the partial trace-based approach. Note that, instead of constituting the local regions with only two useful nodes one of which is the hub, one may also consider a larger subset of functional nodes including the hub, and perform measurements on the rest of the nodes to obtain the regionally localized entanglement. However, this broader approach demands a detailed characterization of multipartite entanglement, which becomes especially challenging in noisy environments. Hence, an intriguing question remains as to whether, in such general setups, meaningful bounds on regionally localized entanglement can still be established, or if they cease to exist.

Summarizing, we focused on regions constituted of two qubits including the hub, and proved that for several classes of multipartite quantum states, the total regionally localized entanglement is upper bounded by a system-size-dependent block entanglement. Our results apply to permutation-symmetric states such as generalized Greenberger Horne Zeilinger  states, Dicke states, and their superpositions, as well as to states within specific magnetization sectors that break permutation symmetry. For fixed-excitation states, especially generalized W states with a single excitation, we derived both lower and upper bounds for regionally localized entanglement, governed by block entanglement. While these bounds differ from those in symmetric states, they converge as the number of excitations increases. Additionally,   our numerical simulations for  Haar-uniformly generated random states revealed that   similar upper and lower bounds on regionally localized entanglement in terms of block entanglement also hold. These findings confirm that the block entanglement serves as a consistent bound across various structured and random multipartite quantum systems.

We further examined the impact of local Markovian and non-Markovian phase-flip noise on each qubit. We demonstrated that, even under moderate decoherence, the upper bounds on regionally localized entanglement are preserved, though the lower bound can be violated when the block entanglement and system-size are small. This establishes the robustness of the upper bounds across a broad range of quantum scenarios.

\acknowledgements

The Authors acknowledge the use of \href{https://github.com/titaschanda/QIClib}{QIClib} -- a modern C++ library for general purpose quantum information processing and quantum computing. A.K.P and A.S.D acknowledge the support from the Anusandhan National Research Foundation (ANRF) of the Department of Science and Technology (DST), India, through the Core Research Grant (CRG) (File No. CRG/2023/001217, Sanction Date 16 May 2024). A.S.D. acknowledges support from the project entitled "Technology Vertical - Quantum Communication'' under the National Quantum Mission of the Department of Science and Technology (DST)  ( Sanction Order No. DST/QTC/NQM/QComm/2024/2 (G)).

\appendix

\section{Negativity}
\label{app:negativity}

For a bipartition $A:B$ of a set of qubits in a state $\rho$, negativity is defined as~\cite{peres1996,horodecki1996,zyczkowski1998,vidal2002,lee2000}
\begin{eqnarray} 
\mathcal{N}(\rho)=\frac{||\rho^{T_{B}}||-1}{2},
\end{eqnarray}
which corresponds to the absolute value of the sum of negative eigenvalues, $\lambda$, of $\rho^{T_{B}}$, given by 
\begin{eqnarray}
\mathcal{N}(\rho)=\left|\sum_{\lambda_i<0}\lambda_i\right|.
\end{eqnarray}
Here, $||\varrho|| = \mbox{Tr}\sqrt{\varrho^{\dagger}\varrho}$ is the trace norm of the density operator $\varrho$, computed as the sum of the absolute values of the eigenvalues of $\varrho$. The matrix $\rho^{T_{B}}$ is obtained by performing partial transposition of the density matrix $\rho$ with respect to the subsystem  $B$. In this paper, we are specifically interested in situations where the subsystem $A$ consists of only one qubit, in which case $0\leq E(\rho)\leq 1/2$. In order for rescaling the value between $0$ and $1$, in this paper, we consider the entanglement $E(\rho)$ over a bipartition $1:\text{rest}$, as quantified by negativity, to be $E(\rho)=2\mathcal{N}(\rho)$. 

\section{Monogamy of entanglement}
\label{app:monogamy}


Consider an $N$-qubit system described by the state $\rho_{12\cdots N}$, where the qubits are labeled as $1,2,\cdots,N$. An $N$-qubit quantum state $\rho_{12\cdots N}$ is said to be \emph{monogamous} with respect to a \emph{bipartite} entanglement measure $E$ and a chosen \textit{nodal qubit} ``$1$" if~\cite{ckw2000,Terhal2004,Christandl2004,Osborne2006,Kim2012a} 
\begin{eqnarray}
    E(\rho)\equiv E(\rho_{1:\text{rest}})\geq  \sum_{i=2}^N F_{i}(\rho_{1i})\equiv F (\rho),
    \label{eq:monogamy_condition}
\end{eqnarray}
where $E(\rho)$ is the bipartite entanglement of $\rho_{12\cdots N}$, as quantified by $E$, between the qubit $1$ and the rest of the qubits together.  On the other hand,  $F_{i}(\rho)$ is the bipartite entanglement quantified by the same entanglement measure $E$ over the qubit-pair $(1,i)$, computed with the reduced density matrix $\rho_{1i}=\text{Tr}_{\mathcal{S}\backslash\{1,i\}}\left[\rho_{12\cdots N}\right]$ obtained by tracing out all other qubits except the qubits $1$ and $i$.  Unless otherwise stated, in this paper, we always fix qubit ``$1$" as the nodal qubit, and refrain from including it in the notation for brevity, and $\mathcal{S}$ is the set of all qubits, with $\mathcal{S}\backslash\{1,i\}$ being the set of all qubits barring $1$ and $i$. Only for this appendix, we use $\rho$ and $\rho_{12\cdots N}$ interchangeably.  
If Eq.~(\ref{eq:monogamy_condition}) is not satisfied, the state is \emph{non-monogamous}, and when Eq.~(\ref{eq:monogamy_condition}) is satisfied by all $N$-qubit entangled states for a given entanglement measure $E$ and $N$, the measure is called monogamous for all states with $N$-qubits. For a fixed bipartite entanglement measure $E$ and the node $i$,  on the $\left(E(\rho),F(\rho)\right)$-plane, all monogamous states lie on or below the line $F(\rho)=E(\rho)$, while all non-monogamous states are found above it. One can define a \emph{monogamy score} 
\begin{eqnarray}
    \Delta(\rho)=E(\rho)-F(\rho),
    \label{eq:monogamy_score}
\end{eqnarray}
corresponding to the $N$-qubit state $\rho$, which is $\geq 0$ ($<0$)  on and below (above) the line $F(\rho)=E(\rho)$. Note that for all $N$-qubit pure states, the maximum value of $E(\rho)=1$, implying the maximum and the minimum values of $\Delta(\rho)$ are respectively $1$ and $0$ (since $F_{i}(\rho_{1i})\leq 1$ $\forall i$) for all monogamous $N$-qubit pure states for any bipartite entanglement measure $E$. The upper bound is saturated by the $N$-qubit GHZ state, given by $\left(\ket{0}^{\otimes N}+\ket{1}^{\otimes N}\right)/\sqrt{2}$, by virtue of $F(\rho)=0$, and the lower bound is saturated by a $(N-1)$-separable state of the form $\ket{\psi}\otimes_{j=1}^{N-2}\ket{\phi_j}$, where $\ket{\psi}$ is a singlet state.

\section{Connection to entanglement of assistance}
\label{app:eoa}

Consider an $M$-qubit system described by the state $\rho$, constituted of a \emph{fully accessible} partition $\mathcal{S}_0$ of $N_0$ qubits, and another partition $\mathcal{S}$ of $N$-qubits with \emph{restricted accesibility}, where we label the qubits in $\mathcal{S}$ as $1,2,\cdots,N$ (see Fig.~\ref{fig:partition}). For an arbitrary pure state $\rho=\ket{\Psi}\bra{\Psi}$ of the system $\mathcal{S}_0\cup\mathcal{S}$, one defines the
entanglement of assistance (EoA), $\mathcal{E}_a(\rho_{\mathcal{S}})$ of $\rho_\mathcal{S}=\text{Tr}_{\mathcal{S}_0}\left[\rho\right]$ using the chosen \emph{convex} entanglement measure $E$ as 
\begin{eqnarray}
    \mathcal{E}_a(\rho_{\mathcal{S}})&=&\max\sum_ip_i E\left(\ket{\psi_i}\bra{\psi_i}\right),    
\end{eqnarray}
where $E$ is computed in the $1:\text{rest}$ bipartition in $\mathcal{S}$, and the maximization is performed over all possible pure state decompositions of $\rho_{\mathcal{S}}$. Restricting to projection measurements on $\mathcal{S}_0$ that leads to only pure states on $\mathcal{S}$, using the definition of $\mathcal{E}_\mathcal{S}(\rho_\mathcal{S})$ put forward in Eq.~(\ref{eq:lqc}),  one may write 
\begin{eqnarray}\label{eq:condition_1}
    \mathcal{E}_{\mathcal{S}}(\rho_\mathcal{S})\leq \mathcal{E}_a(\rho_{\mathcal{S}}),
\end{eqnarray}
since each measurement in $\mathcal{M}_0$ on $\mathcal{S}_0$ leads to an ensemble representing a pure state decomposition of $\rho_{\mathcal{S}}$. Note further that (see Sec.~\ref{sec:definitions})
\begin{eqnarray}
    \rho_{1i}=\text{Tr}_{(\mathcal{S}_0\cup\mathcal{S})\backslash\{1,i\}}\left[\rho\right]=\text{Tr}_{\mathcal{S}\backslash\{1,i\}}\left[\rho_{\mathcal{S}}\right],
\end{eqnarray}
which, along with arguments similar to that for Eq.~(\ref{eq:condition_1}), leads to 
$\mathcal{F}_i(\rho)\leq \mathcal{E}_a(\rho_{1i})$ $\forall i=2,3,\cdots,N$, resulting in 
\begin{eqnarray}\label{eq:condition_2}
    \mathcal{F}_{\mathcal{S}}(\rho)\leq \sum_{i=2}^N\mathcal{E}_a(\rho_{1i}). 
\end{eqnarray}

Refs.~\cite{Gour2005,Gour_2007,Buscemi_2009,Kim_2012,Kim2012a,Kim2014,Kim2018,Guo2018,Yang2019,Jin2019,Jin2019b,shen2024} show that  
\begin{eqnarray}
    \mathcal{E}_a(\rho_{\mathcal{S}})\leq \sum_{i=2}^N\mathcal{E}_a(\rho_{1i}),
    \label{eq:EoA_known_result_1}
\end{eqnarray}
and in the special case of $\mathcal{S}_0=\emptyset$, 
\begin{eqnarray}
    E(\rho)\leq \sum_{i=2}^N\mathcal{E}_a(\rho_{1i}),
    \label{eq:EoA_known_result_2}
\end{eqnarray}
where $E(\rho)$ is computed in terms of von Neumann entropy of the qubit $``1"$ for the $N$-qubit pure state $\rho$. Since $\mathcal{E}_{\mathcal{S}}(\rho)\leq \sum_{i=2}^N\mathcal{E}_a(\rho_{1i})$  by virtue of Eqs.~(\ref{eq:condition_1}) and (\ref{eq:EoA_known_result_1}), Eq.~(\ref{eq:condition_2}) makes the hierarchy of $\mathcal{E}_{\mathcal{S}}$ and $\mathcal{F}_{\mathcal{S}}$ non-trivial.

\bibliography{ref}

\end{document}